\documentclass[10pt]{article}
\usepackage[brazil]{babel}
\usepackage{url}
\usepackage{amssymb,amsxtra}
\usepackage{amsmath,color,graphicx}
\usepackage{lineno}
\usepackage{graphicx}
\RequirePackage{geometry}
\usepackage{indentfirst}

\usepackage{blindtext}
\usepackage{multicol}
\title{Second multicols Demo}
\author{Overleaf}
\date{April 2021}

\begin{document}

\def\chaptername{}
\def\contentsname{Sum\'{a}rio}
\def\listfigurename{Figuras}
\def\listtablename{Tabelas}
\def\abstractname{Resumo}
\def\appendixname{Ap\^{e}ndice}
\def\refname{\large Refer\^{e}ncias bibliogr\'{a}ficas}
\def\bibname{Bibliografia}
\def\indexname{\'{I}ndice remissivo}
\def\figurename{\small Figura~}
\def\tablename{\small Tabela~}
\def\pagename{\small Pag.}
\def\seename{veja}
\def\alsoname{veja tamb\'em}
\def\na{-\kern-.4em\raise.8ex\hbox{{\tt \scriptsize a}}\ }
\def\pa{\slash \kern-.5em\raise.1ex\hbox{p}\ }
\def\ro{-\kern-.4em\raise.8ex\hbox{{\tt \scriptsize o}}\ }
\def\no{n$^{\underline{\rm o}}$}
\setcounter{tocdepth}{3}

\clearpage
\pagenumbering{arabic}

\thispagestyle{empty}
\parskip 8pt

\vspace*{0.2cm}
\begin{center}

\ \\

{\huge \bf Uma breve hist\'{o}ria do \textit{spin}}\\
\vspace*{1.0cm}

{\Large \bf \it Francisco Caruso;$^{\,1}$ Vitor Oguri$^{\,2}$}\\[2.em]

{{$^{1}$ Centro Brasileiro de Pesquisas F\'{\i}sicas, Coordena\c{c}\~{a}o de F\'{\i}sica de Altas Energias, 22290-180, Rio de Janeiro, RJ, Brazil.}}

{{$^{2}$ Universidade do Estado do Rio de Janeiro, Instituto de F\'{\i}sica Armando Dias Tavares, 20550-900, Rio de Janeiro, RJ, Brasil.}}

\vfill
\end{center}

\noindent \textbf{Resumo}

Faz-se uma an\'{a}lise de alguns pontos hist\'{o}ricos envolvidos na consolida\c{c}\~{a}o do conceito te\'{o}rico de \textit{spin}, introduzido originalmente para explicar a estrutura dos espectros at\^{o}micos na aus\^{e}ncia e na presen\c{c}a de campos eletromagn\'{e}ticos, em 1925, pelos f\'{\i}sicos holandeses Samuel Abraham Goudsmit e George Eugene Uhlenbeck. Reitera-se sua relev\^{a}ncia no \^{a}mbito da descri\c{c}\~{a}o qu\^{a}ntica da mat\'{e}ria.

\noindent \textbf{Palavras-chave:} Spin; Mec\^{a}nica Qu\^{a}ntica; F\'{\i}sica At\^{o}mica; Hist\'{o}ria da F\'{\i}sica.

\vspace*{0.7cm}

\noindent \textbf{Abstract}

A brief analysis is made of some historical points involved in the consolidation of the theoretical concept of \textit{spin}, originally introduced to explain the structure of atomic spectra in the absence and presence of electromagnetic fields, in 1925, by Dutch physicists Samuel Abraham Goudsmit and George Eugene Uhlenbeck. Its relevance in the context of the quantum description of matter is reiterated.





\newpage
\setcounter{page}{1}

\twocolumn

\section{Introdu\c{c}\~{a}o}

\begin{flushright}
\begin{minipage}{5.5cm}
\baselineskip=10pt {\small
\textit{Foi h\'{a} pouco mais de cinquenta anos que George Uhlenbeck e eu introduzimos o conceito de \textit{spin} {\rm [ $\ldots$].} Portanto, n\~{a}o \'{e} surpreendente que a maioria dos jovens f\'{\i}sicos n\~{a}o saiba que o \textit{spin} teve que ser introduzido. Eles acham que ele foi revelado no G\^{e}nesis ou talvez postulado por Sir Isaac Newton, o que a maioria dos jovens f\'{\i}sicos considera ser eventos mais ou menos simult\^{a}neos.
}
\smallskip \\
\vspace*{0.35mm}
\hfill Samuel A. Goudsmit \\
\vspace*{0.35mm}
\hfill discurso, em 1976 na \\
\vspace*{0.35mm}
\hfill  \textit{American Physical Society}}
\end{minipage}
\end{flushright}

O surgimento do conceito de \textit{spin} se entrela\-\c{c}a com o da pr\'{o}pria Mec\^{a}nica Qu\^{a}ntica, ambos introduzidos h\'{a} cerca de 100 anos, em 1925. O \textit{spin}, assim como tantas outras grandezas associadas aos sistemas f\'{\i}sicos e incorporadas \`{a} Mec\^{a}nica Qu\^{a}ntica, surge a partir de analogias com grandezas definidas no contexto da Mec\^{a}\-ni\-ca e do Eletromagnetismo Cl\'{a}ssicos.
O resultado pr\'{a}tico \'{e} que as duas teorias que revolucionaram a F\'{\i}sica no s\'{e}culo XX, tanto a Relatividade Restrita quanto a Mec\^{a}nica Qu\^{a}n\-ti\-ca, s\~{a}o dotadas de estruturas formais condiciona\-das por essas teorias cl\'{a}ssicas, por meio de correspond\^{e}ncias estabelecidas para a forma e a express\~{a}o de seus conceitos, grandezas e princ\'{\i}pi\-os. Esse procedimento mascarou e postergou v\'{a}ri\-as quest\~{o}es de cunho interpretativo, e muitas controv\'{e}rsias originadas nessa \'{e}poca ainda perduram.~No entanto, apesar de seu car\'{a}ter n\~{a}o com\-probat\'{o}rio, as analogias s\~{a}o frut\'{\i}feras.~O ca\-so do \textit{spin} n\~{a}o foi exce\c{c}\~{a}o.

Na Mec\^{a}nica Cl\'{a}ssica, o momento angular est\'{a} associado \`{a}s principais caracter\'{\i}sticas do movimento dos planetas sob a a\c{c}\~{a}o do campo central gravitacional do Sol. Esta mesma grandeza desempenha um papel an\'{a}logo na an\'{a}lise dos sistemas at\^{o}micos, nos quais os el\'{e}trons se movem sob a a\c{c}\~{a}o do campo el\'{e}trico central coulombiano dos n\'{u}cleos.~Assim como o planeta pos\-sui um momento angular orbital em rela\c{c}\~{a}o ao Sol e um momento angular pr\'{o}prio, em re\-la\-\c{c}\~{a}o ao seu centro de massa, o el\'{e}tron, em um \'{a}\-to\-mo, al\'{e}m do momento angular orbital em rela\c{c}\~{a}o ao n\'{u}cleo, apresenta tamb\'{e}m um momento angular pr\'{o}prio, mas que, ao contr\'{a}\-rio do caso planet\'{a}rio, n\~{a}o est\'{a} associado a qualquer movimento do el\'{e}tron, sendo chamado de \textit{momento angular intr\'{\i}nseco} ou \textit{spin}.

Apesar da analogia, a denomina\c{c}\~{a}o de momento intr\'{\i}nseco, em vez de \textit{spin}, talvez fosse mais adequada para afastar  qualquer tentativa de associ\'{a}-lo ao movimento de rota\c{c}\~{a}o como se o el\'{e}tron fosse um corpo extenso. Isto porque, at\'{e} o limite experimental de hoje, que corresponde a dist\^{a}ncias da ordem de $10^{-18}$~cm, o el\'{e}tron n\~{a}o possui estrutura, ou seja, n\~{a}o \'{e} uma part\'{\i}cula composta e, portanto, praticamente, n\~{a}o tem dimens\~{a}o. Diferentemente do momento angular orbital, que depende das intera\c{c}\~{o}es do el\'{e}tron com o n\'{u}cleo ou qualquer campo externo, o \textit{spin} n\~{a}o depende das condi\c{c}\~{o}es externas impostas ao el\'{e}tron; \'{e} um atributo intr\'{\i}nseco, como a massa e a carga el\'{e}trica.

A necessidade desse conceito surge na espectroscopia at\^{o}mica, cujos resultados eram explicados com base em modelos at\^{o}micos planet\'{a}rios, levando os proponentes do conceito a utilizarem dessa analogia para estabelece-lo.
O termo \textit{spin} est\'{a} t\~{a}o arraigado no jarg\~{a}o da F\'{\i}si\-ca, que qualquer tentativa para erradic\'{a}-lo seria t\~{a}o in\'{u}til como propor n\~{a}o usar mais o termo \textit{\'{a}tomo} para objetos que hoje s\~{a}o notadamente divis\'{\i}veis.

No dom\'{\i}nio das intera\c{c}\~{o}es fundamentais eletromagn\'{e}ticas, fracas e fortes, cujos processos s\~{a}o descritos por teorias qu\^{a}nticas, al\'{e}m das tr\^{e}s vari\'{a}veis independentes associadas \`{a}s coordenadas espaciais de uma part\'{\i}cula, ou \`{a}s componentes de seu \textit{momentum} linear, diz-se que o el\'{e}tron, ou qualquer outra part\'{\i}cula elementar do Modelo Padr\~{a}o da F\'{\i}sica de Part\'{\i}culas, possui um novo grau de liberdade, associado a uma nova grandeza ou propriedade intr\'{\i}nseca, o \textit{spin}.

De um ponto de vista mais fundamental, tan\-to na Mec\^{a}nica Cl\'{a}ssica como na Qu\^{a}ntica, devido \`{a} hip\'{o}tese da isotropia espacial, segundo a qual as propriedades do espa\c{c}o s\~{a}o as mes\-mas em qualquer dire\c{c}\~{a}o, h\'{a} uma estreita liga\c{c}\~{a}o formal entre o momento angular, seja orbital ou de \textit{spin}, e o grupo das rota\c{c}\~{o}es. A sobreviv\^{e}ncia das designa\c{c}\~{o}es de grandezas baseadas em modelos mec\^{a}nicos resulta da conex\~{a}o, de car\'{a}ter geral, entre essas grandezas e as simetrias associadas a um sistema, que permitem a generaliza\c{c}\~{a}o de v\'{a}rios conceitos origin\'{a}rios da F\'{\i}si\-ca Cl\'{a}ssica, como o \textit{momentum}, o momento an\-gular e a energia.

Pelo fato de o \textit{spin} n\~{a}o estar associado diretamente \`{a}s coordenadas espaciais de uma par\-t\'{\i}\-cu\-la, os m\'{e}todos matem\'{a}ticos que caracterizam e descrevem as intera\c{c}\~{o}es das part\'{\i}culas que envolvem o \textit{spin} no dom\'{\i}nio n\~{a}o relativ\'{\i}stico s\~{a}o, num certo sentido, bem mais simples do que aqueles que descrevem os processos que envolvem grandezas associadas \`{a}s coordenadas espaciais. Em lugar de equa\c{c}\~{o}es diferenciais  definidas em um espa\c{c}o de fun\c{c}\~{o}es de dimens\~{a}o infinita, em muitos casos, principalmente, envolvendo el\'{e}trons, utilizam-se equa\c{c}\~{o}es em espa\-\c{c}os de dimens\~{a}o finita, de dimens\~{a}o 2, envolven\-do matrizes $(2\times 2)$ e $(2 \times 1)$.

Mais recentemente, devido \`{a} facilidade de abordagem matem\'{a}tica -- apesar de mais abstrata -- e o interesse em t\'{o}picos que se tornaram importantes ao final do s\'{e}culo XX, como o fato de duas  ou mais part\'{\i}culas poderem estar de tal forma conectadas, que o estado de uma n\~{a}o possa ser descrito independentemente do estado das outras, mesmo que estejam espacialmente separadas por milh\~{o}es de anos-luz, existe uma tend\^{e}ncia de se iniciar o estudo da Mec\^{a}nica Qu\^{a}ntica, a partir da defini\c{c}\~{a}o e propriedades do \textit{spin} do el\'{e}tron. N\~{a}o cabe aqui um posicionamento sobre esta escolha, mas apenas destacar que a exist\^{e}ncia dessa correla\c{c}\~{a}o exibida por certos tipos de composi\c{c}\~{a}o de estados de sistemas distintos, denominados \textit{emaranhados qu\^{a}nticos}, al\'{e}m de fundamental no processo de investiga\c{c}\~{a}o de diferentes interpreta\c{c}\~{o}es da Mec\^{a}nica Qu\^{a}ntica, \'{e} a base para novas apli\-ca\-\c{c}\~{o}es tecnol\'{o}gicas, como a Computa\c{c}\~{a}o e a Criptografia Qu\^{a}ntica.

No entanto, a necessidade de se atribuir, em um primeiro momento somente ao el\'{e}tron, esse novo grau de liberdade, surge como decorr\^{e}ncia da an\'{a}lise dos fen\^{o}menos at\^{o}micos, e da interpreta\c{c}\~{a}o dos espectros dos metais alcalinos,\footnote{\, S\~{a}o os elementos qu\'{\i}micos do grupo 1 da Tabela Peri\'{o}dica: l\'{\i}tio ({\tt Li}), s\'{o}dio ({\tt Na}), pot\'{a}ssio ({\tt K}), rub\'{\i}dio ({\tt Rb}) e c\'{e}sio ({\tt Cs}), excetuando-se o hidrog\^{e}nio ({\tt H}), e possuem um \'{u}nico el\'{e}tron de val\^{e}ncia.} sob a a\c{c}\~{a}o de campos magn\'{e}ticos.

A hist\'{o}ria do \textit{spin} se inicia com a descoberta da multiplicidade dos termos espectrais devido ao efeito Zeeman an\^{o}malo, no per\'{\i}odo de  1923--1925, da qual, como frequentemente acontece com a cria\c{c}\~{a}o e introdu\c{c}\~{a}o de novos conceitos em  qualquer ramo da ci\^{e}ncia, resultaram aca\-loradas discuss\~{o}es e disputas sobre a prioridade de ideias, segundo os pr\'{o}prios autores do artigo que, pela primeira vez, introduziu o conceito de \textit{spin} do el\'{e}tron no contexto da es\-pectroscopia at\^{o}mica.~As interpreta\c{c}\~{o}es dos espectros observados, como j\'{a} antecipado, eram baseadas nas teorias de Bohr e Sommerfeld e em algumas regras empiricamente estabelecidas por pesquisadores, principalmente, da Alemanha, Su\'{e}cia, Su\'{\i}\c{c}a e Holanda.~N\~{a}o havia ainda uma teoria qu\^{a}ntica baseada em primeiros princ\'{\i}pios ou e\-qua\-\c{c}\~{o}es gerais, sem hip\'{o}teses \textit{ad hoc}.

A primeira vers\~{a}o da Mec\^{a}nica Qu\^{a}ntica -- na sua formula\c{c}\~{a}o matricial --, apresentada por Werner Heisenberg, Max Born e Pascual Jordan, surgiu tamb\'{e}m no mesmo ano da proposta do conceito de \textit{spin} do el\'{e}tron, h\'{a} 100 anos, em 1925, por Samuel  Goudsmit e  George Uhlenbeck. Pouco tempo depois, em 1926, a vers\~{a}o que se tornou mais difundida -- a formula\c{c}\~{a}o ondulat\'{o}ria --, foi publicada por Erwin Schr\"{o}dinger. Ambas as vers\~{o}es n\~{a}o levam em conta a Relatividade Restrita.

\section{\hspace*{-0.5cm} Os prim\'{o}rdios da es\-pec\-troscopia at\^{o}mica}

\begin{flushright}
\begin{minipage}{5.5cm}
\baselineskip=10pt {\small
\textit{O problema do \'{a}tomo seria resolvido se os f\'{\i}sicos aprendessem a compreender a linguagem dos \textit{espectros}. }
\smallskip

\hfill Arnold J.H. Sommerfeld (1919)
}
\end{minipage}
\end{flushright}

\begin{figure}[htbp]
\centerline{\includegraphics[width=7.0cm]{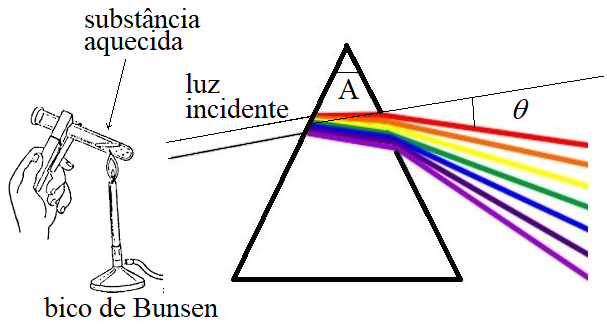}}
\vspace*{-0.3cm}
\caption{\baselineskip=8pt{\small Dispers\~{a}o da luz por um prisma.}}
\label{prisma_dispersa}
\end{figure}

Apesar de os experimentos de espalhamento de feixes de part\'{\i}culas pela mat\'{e}ria, realizados no per\'{\i}odo de 1909--1911,  sob a lideran\c{c}a de Ernest Rutherford, terem sidos determinantes no estabelecimento da estrutura dos \'{a}tomos,
pode-se tra\c{c}ar tamb\'{e}m a origem de muitas das propriedades e ideias sobre a estrutura at\^{o}mica \`{a}s investiga\c{c}\~{o}es realizadas ao final do s\'{e}culo XIX, entre 1855 e 1863, por Robert Bunsen e Gustav Kirchhoff,
 ao observarem a luz emitida por diversas subst\^{a}ncias, associadas a elementos qu\'{\i}\-mi\-cos distintos, quando aquecidas pelo famoso bico de Bunsen. Nessas observa\c{c}\~{o}es, a luz emitida incide sobre um prisma e, ent\~{a}o, \'{e} dispersada de tal modo que as componentes de diferentes cores -- correspondendo a diferentes fre\-qu\^{e}n\-cias, ou comprimentos de onda -- s\~{a}o refra\-ta\-das em diferentes \^{a}ngulos (Figura~\ref{prisma_dispersa}).

\begin{figure}[htbp]
\centerline{\includegraphics[width=7.7cm]{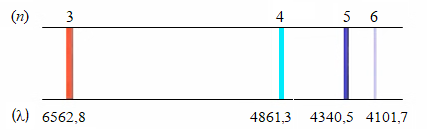}}
\vspace*{-0.3cm}
\caption{\baselineskip=6pt{\small Espectro do \'{a}tomo de hidrog\^{e}nio na faixa vis\'{\i}vel, de frequ\^{e}ncias da ordem de $10^{15}$~Hz. Os comprimentos de onda ($\lambda$) est\~{a}o expressos em angstron (\AA), sendo  1\,\AA\, = $10^{-8}$\,m.}
 }
\label{espectro_h}
\end{figure}

Se os raios difratados s\~{a}o interceptados por um anteparo, como a lente de uma luneta que tem uma escala milimetrada, observa-se, no caso de gases ou vapores contendo, por exemplo, os elementos hidrog\^{e}nio, merc\'{u}rio ou s\'{o}dio, um pa\-dr\~{a}o de linhas luminosas de v\'{a}rias cores, como mostrado na Figura~\ref{espectro_h} para o hidrog\^{e}nio. Cada uma dessas  linhas est\'{a} associada a um comprimento de onda ($\lambda$), ou a uma frequ\^{e}ncia ($\nu$), que pode ser determinado pelo \^{a}n\-gulo ($\theta$) entre a dire\c{c}\~{a}o do feixe de luz incidente no prisma e a dire\c{c}\~{a}o do raio emergente cor\-respondente (Figura~\ref{prisma_dispersa}).\footnote{\, Esse \^{a}ngulo ($\theta$) depende da abertura angular ($A$) do prisma, e do \'{\i}ndice de refra\c{c}\~{a}o associado ao respectivo comprimento de onda.} Esse conjunto de linhas \'{e} chamado \textit{espectro do elemento qu\'{\i}mico},\footnote{\, Em geral as subst\^{a}ncias exibem linhas espectrais, devidas aos \'{a}tomos, e regi\~{o}es onde a radia\c{c}\~{a}o est\'{a} dis\-tri\-bu\'{\i}\-da continuamente -- bandas espectrais -- tendo como origem as mol\'{e}culas.} e cada elemento est\'{a} associado a um conjunto de linhas espectrais caracter\'{\i}stico, como se fosse uma esp\'{e}cie de ``impress\~{a}o digital'', \'{u}nica para cada elemento.


De particular import\^{a}ncia no desen\-vol\-vi\-men\-to da F\'{\i}sica At\^{o}mica foi a caracteriza\c{c}\~{a}o do espectro de linhas mais simples, correspondente tamb\'{e}m ao \'{a}tomo mais simples -- o \'{a}tomo de hidrog\^{e}nio. O espectro do hidrog\^{e}nio (Figura~\ref{espectro_h}) foi observado por  Anders  {\AA}ngstr\"{o}m, 
em 1853, e sua regularidade, matematizada por Johann  Balmer, 
cerca de trinta anos depois, em 1885. Baseando-se nas medidas de {\AA}ngstr\"{o}m, Balmer encontra uma express\~{a}o capaz de reproduzir os comprimentos de onda de cada linha do espectro observado. A f\'{o}rmula foi escrita de forma mais sugestiva por Johannes  Rydberg, em 1888, 
como
\begin{equation} \label{Balmer}
\displaystyle
\frac{1}{\lambda} = R_H \left( \frac{1}{2^2} - \frac{1}{n^2} \right)
 \quad  (n=3,4,5,6)
\end{equation}
sendo o comprimento de onda ($\lambda$) expresso em metros, e $R_H \simeq 1,\!097 \times 10^7$\,m$^{- \scriptstyle 1}$, a constante de Rydberg.\footnote{\, O valor recomendado pelo \textit{Committee on Data for Science and Technology} (CODATA), em 2022, \'{e} $10\, 973\, 731,\! 568\, 157\, (12)$~m$^{- \scriptstyle 1}.$}

Posteriormente, outros conjuntos de s\'{e}ries de linhas espectrais associadas ao hidrog\^{e}nio, em regi\~{o}es n\~{a}o vis\'{\i}veis, do infravermelho e do ultravioleta, foram observadas e descritas por uma generaliza\c{c}\~{a}o da f\'{o}rmula de Balmer. Expressa em termos da frequ\^{e}ncia por Walter Ritz,
em 1908, cada frequ\^{e}ncia do espectro da radia\c{c}\~{a}o emitida, ou termo espectral, \'{e} dada por\footnote{\, Estritamente, a f\'{o}rmula \'{e} v\'{a}lida apenas para propaga\c{c}\~{a}o da luz no v\'{a}cuo.}
\begin{equation} \label{Ritz}
\qquad
\displaystyle
\nu_{mn} = c\, R_H \left( \frac{1}{m^2} - \frac{1}{n^2} \right)
\end{equation}
em que $n > m=1,2,3,4,5$ e $c\simeq3,\!0 \times 10^{8}$~m/s \'{e} a velocidade da luz no v\'{a}cuo,\footnote{\, $c = 299\,792\,458$~m/s  --  CODATA (2022).}  e $\nu_{mn}$ \'{e} a fre\-qu\^{e}n\-cia da radia\c{c}\~{a}o eletromagn\'{e}tica emitida associada a cada linha espectral, em hertz (Hz).



A equa\c{c}\~{a}o~(\ref{Ritz}) \'{e} igualmente v\'{a}lida para os metais alcalinos, com a respectiva constante de Rydberg e,
apesar da generaliza\c{c}\~{a}o, continuava sendo um resultado emp\'{\i}rico, n\~{a}o explicado nem pela Mec\^{a}nica Cl\'{a}ssica nem pelo Eletromagnetismo. A primeira explana\c{c}\~{a}o compat\'{\i}vel com os dados ocorreu somente em 1913, com Niels Bohr (Se\c{c}\~{a}o \ref{bohr}), mas antes, por\'{e}m, deve-se recordar alguns aspectos do efeito Zeeman e da descoberta do el\'{e}tron, ainda no final do s\'{e}culo~XIX.



\section{A g\^{e}nesis -- o ano de 1897: o efeito Zeeman e a descoberta do el\'{e}tron}

\begin{flushright}
\begin{minipage}{5.5cm}
\baselineskip=10pt {\small
\textit{A natureza nos presenteia com surpresas, incluindo a do Prof.~Lorentz. Rapidamente descobrimos que existem muitas exce\c{c}\~{o}es \`{a} regra de dividir as raias apenas em tripletos. }
\smallskip

\hfill Pieter Zeeman
}
\end{minipage}
\end{flushright}
\vspace*{0.5cm}


Desde 1886, Pieter Zeeman, \`{a} \'{e}poca assistente de Hendrik  Lorentz, vinha tentando, sem sucesso, detectar a influ\^{e}ncia do campo magn\'{e}\-ti\-co no espectro do s\'{o}dio ({\tt Na}), mes\-mo ap\'{o}s as tentativas malogradas de Faraday, em 1862, de observar a altera\c{c}\~{a}o do espectro luminoso por um campo magn\'{e}tico. Sabendo-se que a luz, de qualquer origem, n\~{a}o interage com o campo magn\'{e}tico, \'{e} preciso supor que este campo atue sobre o movimento das part\'{\i}culas carregadas, no interior da mat\'{e}ria, respons\'{a}veis pela emiss\~{a}o, alterando sua frequ\^{e}ncia e consequentemente a da luz do espectro. Com essa espectativa, finalmente, em 1897~\cite{Zeeman_2}, j\'{a} utilizando redes de difra\c{c}\~{a}o de Rowland com cerca de 600~linhas/mm no lugar de prismas, e campos magn\'{e}ticos mais intensos -- da ordem de 1\,tesla (T) -- obtidos com bobinas de Ruhmkorff, observa o alargamento de uma linha espectral do s\'{o}dio ({\tt Na}) e, posteriormente, o desdobramento de uma linha do c\'{a}dmio ({\tt Cd}), sob a a\c{c}\~{a}o do campo.

A primeira tentativa para elucidar esse tipo de obser\-va\-\c{c}\~{a}o de Zeeman, baseou-se na hip\'{o}tese de Lorentz de que todos os corpos s\~{a}o constitu\'{\i}dos por part\'{\i}culas eletricamente carregadas -- \textit{os \'{\i}ons} --,\footnote{\, Para Lorentz, existiriam \'{\i}ons com carga positiva e \'{\i}ons com carga negativa, tal que em um diel\'{e}trico, os \'{\i}ons negativos poderiam oscilar em rela\c{c}\~{a}o \`{a}s suas posi\c{c}\~{o}es de equil\'{\i}brio, e todos os processos eletromagn\'{e}ticos de intera\c{c}\~{a}o da luz com a mat\'{e}ria poderiam ser analisados a partir dessa hip\'{o}tese, a qual ele j\'{a} havia utilizado para a explica\c{c}\~{a}o da dispers\~{a}o da luz por diel\'{e}tricos.} e que a luz emitida pelos elementos tem sua origem no movimento peri\'{o}dico desses ``\'{\i}ons'', ou seja, que a frequ\^{e}ncia da luz emitida seria igual a frequ\^{e}ncia ($\omega$) do movimento da part\'{\i}cula. Durante muitos anos, a maioria (para n\~{a}o dizer a totalidade) dos f\'{\i}sicos endossou esse erro, at\'{e} o esclarecimento oferecido por Bohr. 

Fundamentando-se na Mec\^{a}nica Cl\'{a}ssica, e considerando que sob a a\c{c}\~{a}o de um campo mag\-n\'{e}\-ti\-co ($\vec B$) um ``\'{\i}on'' de carga $q$ e velocidade orbital $\vec v$ sofreria a a\c{c}\~{a}o de uma for\c{c}a $\vec F = q \, \vec v \times \vec B$,\footnote{\, Expressa no Sistema Internacional de Unidades (SI). Essa for\c{c}a, posteriormente, foi chamada de \textit{for\c{c}a de Lorentz}.} Lorentz 
estima que a varia\c{c}\~{a}o da frequ\^{e}ncia do movimento ($\Delta \omega$) seria dada por
$$ \Delta \omega = \pm \frac{q}{2m} \, B $$
conforme o sentido do movimento, sendo $m$ a massa da part\'{\i}cula.\footnote{\, Esse resultado pressupunha que o campo magn\'{e}tico fosse perpendicular ao plano do movimento da part\'{\i}cula.}

A partir do desdobramento das linhas espectrais,\footnote{\, A observa\c{c}\~{a}o do desdobramento depende da di\-re\-\c{c}\~{a}o, em rela\c{c}\~{a}o ao campo magn\'{e}tico, segundo a qual se observa a luz emitida. Se a linha de visada \'{e} perpendicular ao campo observa-se o desdobramento em tr\^{e}s novas linhas espectrais e, se paralela ao campo, observa-se o aparecimento em apenas duas linhas.} correspondentes a frequ\^{e}ncias da ordem de $10^{11}$~Hz, em concord\^{a}ncia com as estimativas de Lorentz, al\'{e}m de inferir que as part\'{\i}culas respons\'{a}veis pela emiss\~{a}o da luz tinham carga negativa, Zeeman estimou a raz\~{a}o carga/massa ($q/m$) como da ordem de $10^{8}$~C/g (coulomb/gra\-ma). Sendo cerca de 1\,800 vezes maior que raz\~{a}o carga/mas\-sa dos \'{\i}ons de hidrog\^{e}nio, determinada pela eletr\'{o}lise.

\begin{figure}[htbp]
\centerline{\includegraphics[height=7.0cm]{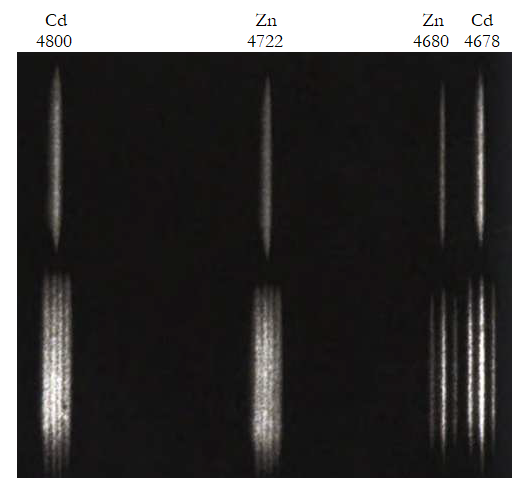}}
\vspace*{-0.1cm}
\caption{\baselineskip=8pt{\small As primeiras imagens dos desdobramentos das linhas espectrais, do zinco ({\tt Zn}) e do c\'{a}dmio ({\tt Cd}), apresentadas por Preston~\cite{Preston}. Sem a presen\c{c}a do campo (parte superior) e sob a a\c{c}\~{a}o do campo magn\'{e}tico (parte inferior). As linhas sem a presen\c{c}a do campo s\~{a}o identificadas pelos comprimentos de onda da luz emitida, expressos em angstrons.}}
\label{preston_zn_cd}
\end{figure}

Apesar das observa\c{c}\~{o}es e das estimativas pioneiras de Zeeman, as primeiras fotografias do desdobramento das linhas espectrais foram  obtidas por Thomas Preston~\cite{Preston}, em Dublin, tam\-b\'{e}m ao final do ano de 1897, ao refazer experimentos an\'{a}logos ao de Zeeman (Figura~\ref{preston_zn_cd}) com um campo magn\'{e}tico mais intenso, da ordem de  2,0\,T, obtendo um valor compat\'{\i}vel com o de Zeeman para a rela\c{c}\~{a}o carga/massa.

No entanto, conforme exemplificado na Figu\-ra~\ref{preston_zn_cd}, ao passo que as observa\c{c}\~{o}es de Preston com re\-la\-\c{c}\~{a}o \`{a}s linhas associadas aos comprimentos de onda, na faixa do azul, de 4\,680\,\AA\  para o zinco, e de 4\,678\,\AA\ para o c\'{a}dmio, concordavam com as observa\c{c}\~{o}es de Zeeman, ao exibir um desdobramento em tr\^{e}s linhas, aquelas correspondentes a 4\,722\,\AA\  para o zinco, e a 4\,800\,\AA\ para o c\'{a}dmio, desdobravam-se em mais do que tr\^{e}s linhas, n\~{a}o concordando com as previs\~{o}es calcadas na teoria de Lorentz.

Os resultados que se manifestavam como duas ou tr\^{e}s linhas -- os \textit{dubletos} e \textit{tripletos} --  tornaram-se conhecidos  como \textit{efeito  Zeeman normal}, e os novos resultados revelados por Preston, que apresentavam mais do que tr\^{e}s linhas -- os \textit{multipletos} --, receberam o nome de \textit{efeito Zeeman an\^{o}malo}, apesar de se revelarem mais frequentes que os dubletos e tripletos de Lorentz. A manifesta\c{c}\~{a}o desses multipletos s\'{o} seria esclarecida com o conceito de \textit{spin} do el\'{e}tron.


No mesmo ano de 1897, Joseph Larmor~\cite{Larmor} apresenta uma  nova interpreta\c{c}\~{a}o para o efeito Zeeman. Segundo Larmor, o efeito de um campo magn\'{e}tico ($B$) seria o de superpor  \`{a} oscila\c{c}\~{a}o natural do ``\'{\i}on de Lorentz''  a  frequ\^{e}ncia dada por  $ (q/2m)\, B$, como consequ\^{e}ncia das oscila\c{c}\~{o}es das componentes do momento angular da par\-t\'{\i}\-cu\-la  perpendiculares ao campo, independentemente da orienta\c{c}\~{a}o do plano de movimento em rela\c{c}\~{a}o ao campo.
Essa frequ\^{e}ncia passa a ser denominada \textit{frequ\^{e}ncia de Larmor} ($\omega_L$), e as oscila\c{c}\~{o}es das componentes do momento angular, \textit{precess\~{a}o de Larmor} (Ap\^{e}ndice~\ref{Larmor_classica}).

Enquanto a medida da raz\~{a}o carga/massa, 
a partir do efeito Zeeman normal, foi o primeiro ind\'{\i}cio da exist\^{e}ncia de uma das part\'{\i}culas elementares do Modelo Padr\~{a}o da F\'{\i}sica de Par\-t\'{\i}\-culas -- o \textit{el\'{e}tron} --, a descoberta do efeito Zeeman an\^{o}malo por Preston foi a primeira evi\-d\^{e}n\-cia de manifesta\c{c}\~{a}o do \textit{spin} do el\'{e}tron.

A outra evid\^{e}ncia da exist\^{e}ncia do el\'{e}tron, a qual \'{e} creditada como sua descoberta, foi estabelecida tamb\'{e}m em 1897. Joseph Thomson, no laborat\'{o}rio Cavendish de Cambridge, 
j\'{a} conhecendo que os raios cat\'{o}dicos eram feixes de part\'{\i}culas negativamente carregadas, a partir da deflex\~{a}o desses feixes sob a a\c{c}\~{a}o combinada de campos el\'{e}tricos e magn\'{e}ticos, determinou que a raz\~{a}o carga-massa das part\'{\i}culas constituintes dos raios cat\'{o}dico era da mesma ordem de grandeza do valor obtido por Zeeman, independentemente da natureza do  catodo. Essas part\'{\i}culas tamb\'{e}m adquiriam velocidades da ordem de 1/4 da velocidade da luz no v\'{a}cuo, muitos maiores do que a adquirida por qualquer outro corpo, indicando que suas massas eram bem menores do que a dos \'{\i}ons de hidrog\^{e}nio.

A partir dos resultados dos experimentos de Zeeman e Thomson, e dos trabalhos te\'{o}ricos de Larmor, os ``\'{\i}ons negativos'' de Lorentz vieram a ser identificados como el\'{e}trons \cite{Arab}.

\section{O \'{a}tomo f\'{\i}sico de Rutherford e Bohr} \label{bohr}

\begin{flushright}
\begin{minipage}{5.5cm}
\baselineskip=10pt {\small
\textit{Deve haver qualquer coisa por tr\'{a}s de tu\-do isso {\rm [do modelo de Bohr]}. N\~{a}o acredito que o valor da constante de Rydberg possa ser obtido corretamente por acaso.}
\smallskip

\hfill Albert Einstein
}
\end{minipage}
\end{flushright}

Ao final do s\'{e}culo XIX, e in\'{\i}cio do s\'{e}culo XX, apesar dos avan\c{c}os nos estudos sobre o espectro dos elementos  qu\'{\i}micos, e  da consolida\c{c}\~{a}o da hip\'{o}tese at\^{o}mica da mat\'{e}ria, pouco se sabia sobre a estrutura dos \'{a}tomos, ou seja, ainda n\~{a}o havia um modelo at\^{o}mico f\'{\i}sico.~Ap\'{o}s a descoberta do el\'{e}tron, o pr\'{o}prio Thomson
prop\~{o}e um modelo no qual o \'{a}tomo era constitu\'{\i}do de el\'{e}trons em movimento, em um substrato com carga el\'{e}trica positiva uniformemente distribu\-\'{\i}da em um volume esf\'{e}rico.
No entanto, confrontado em uma s\'{e}rie de experimentos realizados por Hans Geiger e Ernest Marsden, nos laborat\'{o}rios de Manchester,
sob a supervis\~{a}o de Ernest Rutherford, n\~{a}o se mostrou compat\'{\i}vel com os resultados.
Esses experimentos consistiram no espalhamento de feixes de part\'{\i}culas $\alpha$ por folhas finas de v\'{a}rios materiais, como o ouro ({\tt Au}) e a prata ({\tt Ag}),
e foram determinantes para o embasamento do conceito f\'{\i}sico de \'{a}tomo.

A grande contradi\c{c}\~{a}o revelada pelos resultados experimentais de Geiger e Marsden, com as previs\~{o}es baseadas no modelo de Thomson, foram as frequentes ocorr\^{e}ncias de grandes desvios dos feixes em rela\c{c}\~{a}o \`{a} dire\c{c}\~{a}o de incid\^{e}ncia. Em 1911, Rutherford~\cite{Rutherford_11},   a partir da an\'{a}lise da distribui\c{c}\~{a}o angular desses espalhamentos, estabelece que os resultados derivavam da hip\'{o}tese de que os \'{a}tomos eram sistemas constitu\'{\i}dos por uma distribui\c{c}\~{a}o de carga $Ze$, praticamente pontual,\footnote{\, Considerando o raio do \'{a}tomo da ordem de $10^{-10}$\,m, Rutherford estima que a dimens\~{a}o dessa pequena regi\~{a}o seria menor que $10^{-14}$\,m. Sabe-se hoje que \'{e} da ordem de $10^{-15}$\,m, ou seja, da ordem de $100\,000$ vezes menor que o \'{a}tomo, aproximadamente, na mesma propor\c{c}\~{a}o de um gr\~{a}o de arroz e um campo de futebol.} rodeada por um n\'{u}mero $Z$ de el\'{e}trons de modo a anular a carga central. Estava descoberta a estrutura nuclear do \'{a}tomo.

O pr\'{o}ximo passo, que permite relacionar os resultados da espectroscopia com a estrutura a\-t\^{o}\-mi\-ca, \'{e} dado por  Niels Bohr~\cite{Bohr}, em 1913, ap\'{o}s passar o ano de 1912 com o grupo de Rutherford, em seu laborat\'{o}rio em Manchester. Mesmo ap\'{o}s ter levado \`{a} descoberta do n\'{u}cleo at\^{o}mico e \`{a} proposta de um modelo para o \'{a}tomo, o programa de pesquisa de Rutherford, ou seja, do grupo de Manchester, n\~{a}o envolvia a espectroscopia at\^{o}mica, nem a preocupa\c{c}\~{a}o com a estabilidade dos modelos at\^{o}micos; estava voltado para o fen\^{o}meno da radioatividade; para o estudo dos n\'{u}cleos dos \'{a}tomos.

Bohr, afastando-se deste programa e, aparentemente, sem preocupa\c{c}\~{a}o com a espectroscopia, tenta encontrar uma solu\c{c}\~{a}o para a suposta instabilidade de um modelo at\^{o}mico do tipo planet\'{a}rio, proposto por Rutherford. Segundo Larmor~\cite{Larmor}, o movimento de uma part\'{\i}cu\-la com carga el\'{e}trica, como o el\'{e}tron, ao redor de um centro de for\c{c}as seria inst\'{a}vel, pois a part\'{\i}cula perderia energia emitindo radia\c{c}\~{a}o eletromagn\'{e}tica.\footnote{\, Apesar do \'{a}tomo ser um sistema constitu\'{\i}do por um n\'{u}cleo eletricamente carregado e el\'{e}trons, e n\~{a}o apenas de el\'{e}trons.}

Por essa \'{e}poca, desde 1900, j\'{a} era do conhecimento a hip\'{o}tese de Max Planck, 
de que, no fen\^{o}meno da Radia\c{c}\~{a}o de Corpo Negro, a troca de energia entre os constituintes de um corpo e cada componente monocrom\'{a}tica da radia\c{c}\~{a}o, emitida ou absorvida, s\'{o} poderia ser um m\'{u}ltiplo da frequ\^{e}ncia ($\nu$), sendo o valor m\'{\i}nimo dessa  troca de energia igual a $h \nu$, e a constante que caracteriza a chamada \textit{quantiza\c{c}\~{a}o da energia},
$$ h = 6,\!626\,070\,15 \times  10^{-34}\,{\rm J}\! \cdot \!{\rm s} $$
denominada \textit{constante de Planck}.\footnote{\,  Sendo uma das constantes fundamentais da F\'{\i}sica, segundo a \'{u}ltima reuni\~{a}o do CODATA (2022), a constante de Planck ($h$), a velocidade da luz no v\'{a}cuo ($c$), a constante de Boltzmann ($k$), a carga elementar ($e$) e a acelera\c{c}\~{a}o padr\~{a}o da gravidade ($g_n$) s\~{a}o constantes definidas, portanto,  sem incertezas associadas.}

Admitindo que o \'{a}tomo n\~{a}o irradia espontaneamente, utilizando-se da rela\c{c}\~{a}o entre a energia e a frequ\^{e}ncia estabelecida pela Mec\^{a}ni\-ca Cl\'{a}ssica para o movimento em \'{o}rbita el\'{\i}ptica de uma part\'{\i}cula, de massa $m$ e carga el\'{e}trica negativa ($- e$), sob a a\c{c}\~{a}o de uma for\c{c}a central el\'{e}trica coulombiana, e de um outro argumento arbitr\'{a}rio,\footnote{\, \textit{A posteriori}, tudo indica que, possivelmente, esse argumento, a partir do qual Bohr introduz a quantiza\c{c}\~{a}o da energia no \'{a}tomo de hidrog\^{e}nio, baseou-se  na f\'{o}rmula de Balmer, ap\'{o}s ele tomar ci\^{e}ncia dos trabalhos de Rydberg. Dessa  considera\c{c}\~{a}o, entretanto, decorre a conclus\~{a}o original de que a frequ\^{e}ncia da luz emitida ou absorvida por um \'{a}tomo \textit{n\~{a}o era} a frequ\^{e}ncia do movimento do el\'{e}tron, hip\'{o}tese que vai de encontro \`{a}s hi\-p\'{o}teses anteriores sobre os modelos at\^{o}micos. Este foi um passo decisivo para a futura compreens\~{a}o do \'{a}tomo.}  sobre a rela\c{c}\~{a}o entre a energia, a frequ\^{e}ncia do movimento e a constante de Planck, Bohr estabelece que:
\begin{description}
\item[ ] (i) um \'{a}tomo pode existir em determinados \textit{estados es\-ta\-cion\'{a}rios} (\'{o}rbitas est\'{a}veis), associados a n\'{\i}veis discretos de energia, cujos valores, para o \'{a}tomo de hidrog\^{e}nio, s\~{a}o dados por
\begin{eqnarray*}
   E_n &=&  - \displaystyle \left(\frac{m_e\,e^4}{8\,\varepsilon_\circ^2\,h^2} \right) \, \left( \frac{1}{n^2}\right) \ ({\rm SI}) \\
   &=& - 13,\!6 \, \left( \frac{1}{n^2}\right) \ ({\rm eV}) 
\end{eqnarray*}
sendo $n=1,2,3,\ldots$, $m_e\! \simeq\! 9,\!11 \times 10^{-31}$\,kg a massa do el\'{e}tron, e $\varepsilon_\circ\! = \!8,\!85\times 10^{-12}$\,F/m, a permissividade el\'{e}trica do v\'{a}cuo;\footnote{\, $m_e = 9,\!109\,383\,713\,9\,(28) \times 10^{-31}$\,kg e $\varepsilon_\circ = 8,\!854\,187\,818\,8\,(14) \times 10^{-12}$\,F/m -- CODATA (2022).}

\item[ ] (ii) a emiss\~{a}o (ou absor\c{c}\~{a}o) de radia\c{c}\~{a}o eletromagn\'{e}tica s\'{o} ocorre durante a transi\c{c}\~{a}o entre dois estados es\-ta\-cio\-n\'{a}rios, de n\'{\i}veis de energias $E_m$ e $E_n$, sendo a frequ\^{e}ncia ($\nu$) da radia\c{c}\~{a}o emitida (ou absorvida)  da\-da por
$$ \nu = \frac{\big| E_m - E_n \big|}{h}.
$$
\end{description}

Comparando a previs\~{a}o da frequ\^{e}ncia da luz emitida por um \'{a}tomo de hidrog\^{e}nio com a f\'{o}r\-mu\-la generalizada de Balmer, Bohr  determina que a constante de Rydberg pode ser calculada por
$$ R_H = \frac{m_e\,e^4}{8\,\epsilon_\circ^2\,h^3\, c} \simeq 1,\!09 \times  10^7~{\rm m}^{-\scriptstyle 1} $$

O modelo at\^{o}mico de Bohr teve um grande impacto na espectroscopia. Nas palavras de Einstein:

\vspace*{-0.4cm}
\begin{quotation}
\noindent \baselineskip=10pt {\small\textit{Ent\~{a}o a frequ\^{e}ncia da luz n\~{a}o depende em nada da frequ\^{e}ncia [do movimento] do el\'{e}tron \ldots Isto \'{e} uma con\-quista e\-nor\-me!}}
\end{quotation}

\vspace*{-0.4cm}
Sua express\~{a}o para a energia \'{e} v\'{a}lida para uma ampla classe de \'{a}tomos; os chamados \'{a}\-to\-mos hidrogen\'{o}ides, aqueles com um \'{u}nico el\'{e}\-tron de val\^{e}ncia, como os metais alcalinos (grupo I), os \'{\i}ons de metais alcalinoterrosos (grupo II), --  {\tt Ca}$^+$,  {\tt Sr}$^+$, $\ldots$,-- e \'{\i}ons com o {\tt He}$^+$, {\tt Li}$^{++}$ e {\tt Be}$^{+++}$, e quaisquer de seus is\'{o}topos.\footnote{\, Nesses casos, faz-se a substitui\c{c}\~{a}o $ e^2 \, \rightarrow \, Z e^2$, em que $Z$ \'{e} o n\'{u}mero at\^{o}mico, na constante de Rydberg, ou na express\^{a}o da energia.}

A partir de ent\~{a}o, passa a haver uma in\-ten\-sa atividade na espectroscopia at\^{o}mica. A pro\-cu\-ra de um melhor entendimento da estru\-tu\-ra at\^{o}mica vai levar, por fim, \`{a} necessidade de se propor o \textit{spin}.
\vspace*{0.5cm}


\section{A compreens\~{a}o da estrutura fina das linhas espectrais por Som\-merfeld e a experi\^{e}ncia de Stern-Ger\-lach}

\begin{flushright}
\begin{minipage}{5.5cm}
\baselineskip=10pt {\small
\textit{Todas as leis integrais das linhas espectrais e da teoria at\^{o}mica emanam originalmente da teoria qu\^{a}ntica. \'{E} o misterioso \'{o}rg\~{a}o no qual a Natureza toca sua m\'{u}sica dos espectros e, de acordo com seu ritmo, regula a estrutura dos \'{a}tomos e n\'{u}cleos.}
\smallskip

\hfill Arnold Sommerfeld
}
\end{minipage}
\end{flushright}

Apesar de ter relacionado o momento angular do el\'{e}tron \`{a} constante de Planck, a incorpora\c{c}\~{a}o efetiva do momento angular no modelo de Bohr com \'{o}rbitas el\'{\i}pticas foi implementada por
Sommerfeld~\cite{Sommerfeld_l, Sakurai_b, Caruso-Oguri_FM}, ao final de 1915 e no in\'{\i}cio de 1916. Considerando que a energia ci\-n\'{e}tica do el\'{e}tron fosse dada pela express\~{a}o relativ\'{\i}stica de Einstein, e impondo regras de quan\-ti\-za\c{c}\~{a}o associadas \`{a}s componentes radiais e azimutais, correspondentes  ao  \textit{momentum} e ao momento angular, obt\'{e}m a express\~{a}o para a energia em fun\c{c}\~{a}o de n\'{u}meros inteiros n\~{a}o ne\-ga\-ti\-vos, como
\begin{equation}\label{Sommer_1}
   \displaystyle
E_{n_r, n_\varphi}\! =\! E_\circ \, \left[\! 1\ + \ \left( \frac{\alpha}{n_r + \sqrt{n_\varphi^2 - \alpha^2}} \right)^2 \! \right]^{-1/2}    
\end{equation}
\noindent
em que
$$\displaystyle \left\{ \begin{array}{l}
                                                               n_r = 0,1,2, \ldots , (n-1) \\
                                                              n_\varphi = 1,2,3,\ldots , n \\
                                                              n_r + n_\varphi = n =1,2,3,\ldots
                                                              \end{array}
                                                   \right. ,$$
$E_\circ = m_e c^2 \simeq 0,\!511 \times 10^6$\,eV \'{e} a energia de repouso do el\'{e}tron, $n= (n_r+n_\varphi)$ \'{e} o n\'{u}mero qu\^{a}ntico principal, $n_r$ -- n\'{u}mero qu\^{a}ntico radial --, devido \`{a} quantiza\c{c}\~{a}o da  componente radial do \textit{momentum}, $n_\varphi$ -- n\'{u}mero qu\^{a}ntico secund\'{a}rio --, associado \`{a} quantiza\c{c}\~{a}o do momento angular,  e $\displaystyle \alpha = \left( \frac{1}{4 \pi \varepsilon_\circ} \right) \, \frac{e^2}{\hbar c} \simeq  \frac{1}{137}$ \'{e} a \textit{constante de estrutura fina} de Sommerfeld.\footnote{\, $\hbar = h/2\pi = 1,\!054\,571\,817 \times  10^{-34}\,{\rm J}\! \cdot \!{\rm s}$ \, e \, $\displaystyle \alpha^{-1} = 137,\!035\,999\,177\,(21)$ -- CODATA (2022).}


Devido \`{a} ordem de grandeza de $\alpha$, a energia pode ser expressa, aproximadamente, por
$$
 \displaystyle
E_{n,n_\varphi} = E_\circ - \frac{13,\!6}{n^2} \, \left[ 1 \ + \ \frac{\alpha^2}{n^2} \,\left( \frac{n}{n_\varphi} - \frac{3}{4} \right) \right]  \ {\rm (eV)}
$$

A f\'{o}rmula de Sommerfeld, equa\c{c}\~{a}o~(\ref{Sommer_1}), im\-pli\-ca algumas consequ\^{e}ncias not\'{a}veis, dentre as quais:

\vspace*{-0.5cm}
\begin{description}
\item[ ] (i) o desdobramento das linhas do espectro dos \'{a}tomos, em rela\c{c}\~{a}o \`{a}s linhas determinadas pela f\'{o}rmula de Bohr,  mesmo na aus\^{e}ncia de um campo magn\'{e}tico externo, ou seja, independentemente do efeito Zeeman;

\item[ ] (ii) os valores da energia, extraordinariamente, s\~{a}o iguais \`{a}queles calculados pela expres\-s\~{a}o obtida anos depois, em 1928, 
a partir da equa\c{c}\~{a}o relativ\'{\i}stica de Dirac, a qual j\'{a} incorpora o \textit{spin} do el\'{e}tron.
\end{description}

Deve ser lembrado que, em 1916, ainda n\~{a}o havia nem o conceito de \textit{spin}, nem a equa\c{c}\~{a}o de Dirac.~Algumas hip\'{o}teses sobre essa coinci\-d\^{e}ncia, ou casualidade, s\~{a}o apresentadas por Ya\-kov I. Granovskii~\cite{Granovskii}.

Com respeito ao primeiro item, o pequeno desdobramento, que segundo a f\'{o}rmula de Sommerfeld seria uma corre\c{c}\~{a}o da ordem de $10^{-4}$\,eV, correspondente a uma frequ\^{e}ncia de $10^{11}$\,Hz,  revelando a chamada \textit{estrutura fina} do espectro, j\'{a} tinha sido reportado em outro contexto por Albert Michelson e Edward Morley, 
em 1887, ao observarem a linha vermelha 
do espectro do hidrog\^{e}nio por interfer\^{e}ncia luminosa, mas s\'{o} se revelou aos espectroscopistas ap\'{o}s se manifestar nas linhas do s\'{o}dio.  A revela\c{c}\~{a}o da estrutura fina foi  mais uma manifesta\c{c}\~{a}o do \textit{spin} do el\'{e}tron, agora na aus\^{e}ncia de um campo mag\-n\'{e}\-ti\-co.

Apesar de ter surgido na elabora\c{c}\~{a}o de um modelo at\^{o}mico, o papel da constante de estrutura fina ampliou-se com o desenvolvimento da Eletrodin\^{a}mica Qu\^{a}ntica (QED);  de uma constante que caracteriza um fen\^{o}meno es\-pec\-tros\-c\'{o}\-pi\-co para um ``par\^{a}metro de acoplamento'' que caracteriza as intera\c{c}\~{o}es entre el\'{e}trons e f\'{o}tons. Por raz\~{o}es hist\'{o}ricas, ainda \'{e} conhecida como constante de estrutura fina; no entanto,  1/137 \'{e} o valor assint\'{o}tico quando a energia de intera\c{c}\~{a}o el\'{e}tron-f\'{o}ton \'{e} bem menor que a energia de repouso do el\'{e}tron~\cite{Sakurai_a, Halzen, Caruso_QCD}.
%

Ainda em 1916, Sommerfeld~\cite{Sommerfeld_m} 
prop\~{o}e que na presen\c{c}a de um campo magn\'{e}tico, o desdobramento das linha espectrais era caracterizado por um novo n\'{u}mero qu\^{a}ntico ($m$) -- n\'{u}mero qu\^{a}ntico azimutal ou magn\'{e}tico --,  associado \`{a} inclina\c{c}\~{a}o das \'{o}rbitas eletr\^{o}nicas em rela\c{c}\~{a}o \`{a} dire\c{c}\~{a}o do campo, tal que o cosseno do \^{a}ngulo polar $\theta$ entre as dire\c{c}\~{o}es do campo e a normal ao plano da \'{o}rbita ou, equivalentemente, com o momento angular ou com o momento magn\'{e}tico,\footnote{\, De acordo com o Eletromagnetismo, o momento magn\'{e}tico de uma part\'{\i}cula com carga el\'{e}trica \'{e} proporcional ao seu momento angular.} era dado por $\cos \theta = m/n_\varphi$. Uma vez que $n_\varphi = 1,2,3,\ldots$ e $|m| \leq n_\varphi$  fossem inteiros implicava que a inclina\c{c}\~{a}o dos planos das  \'{o}rbitas variava discretamente, comportamento que ficou conhecido como  \textit{quantiza\c{c}\~{a}o espacial}. Para Sommerfeld, os estados estacion\'{a}rios de um \'{a}tomo seriam caracterizado por tr\^{e}s n\'{u}meros qu\^{a}nticos  ($n, n_\varphi, m$), e as linhas espectrais decorriam de transi\c{c}\~{o}es entre esses estados, ou seja, dependiam das varia\c{c}\~{o}es e combina\c{c}\~{o}es desses n\'{u}\-me\-ros.

A hip\'{o}tese da quantiza\c{c}\~{a}o espacial de Sommerfeld foi verificada por Otto Stern e Walther Gerlach~\cite{Stern-Gerlach}, em 1922, 
no celebrado experimento ao qual muitos  atribuem a primeira evid\^{e}ncia do \textit{spin}. No entanto, a manifesta\c{c}\~{a}o do \textit{spin}, tamb\'{e}m implicitamente, j\'{a} havia ocorrido anteriormente, em 1897, nos experimentos realizados por Thomas Preston, nos quais, como j\'{a} mencionado, observou o efeito Zeeman an\^{o}malo.

De acordo com o Eletromagnetismo, o momento  magn\'{e}tico ($\vec \mu$) de um el\'{e}tron com momento angular orbital $\vec L$ \'{e} igual a
\begin{equation}\label{mu}
\displaystyle \vec \mu \, = \,-  \frac{e}{2m_e}\, \vec L
\end{equation}
Sob a\c{c}\~{a}o de um campo magn\'{e}tico ($\vec B$), a energia potencial ($V$) de intera\c{c}\~{a}o magn\'{e}tica \'{e} dada por
$$ V\! =\!  - \vec \mu \cdot \vec B \! =\!  \left( \!\frac{e}{2m_e} \!\right) \vec L \cdot \vec B \, = \, \left(\! \frac{e}{2m_e} \! \right) L B \cos \theta$$
em que $\theta$ \'{e} o \^{a}ngulo entre o momento angular e o campo.
Se o campo variar rapidamente em uma dire\c{c}\~{a}o $z$ -- havendo um forte gradiente de campo --, a part\'{\i}cula estar\'{a} sob a a\c{c}\~{a}o de uma for\c{c}a ($F_z$) nessa dire\c{c}\~{a}o, proporcional \`{a} componente do momento angular tamb\'{e}m nessa dire\c{c}\~{a}o, ou seja, proporcional ao cosseno de $\theta$, dada por
 $$\displaystyle  F_z \, =\, - \left(\frac{e L}{2m_e}\right) \left(\frac{\partial B}{\partial z}\right) \, \cos \theta$$
\indent Assim, segundo a hip\'{o}tese de quantiza\c{c}\~{a}o espacial de Sommerfeld, um el\'{e}tron em movimento em uma dire\c{c}\~{a}o perpendicular ao campo seria desviado apenas em certas dire\c{c}\~{o}es.

\begin{figure}[htbp]
\centerline{\includegraphics[width=7.5cm]{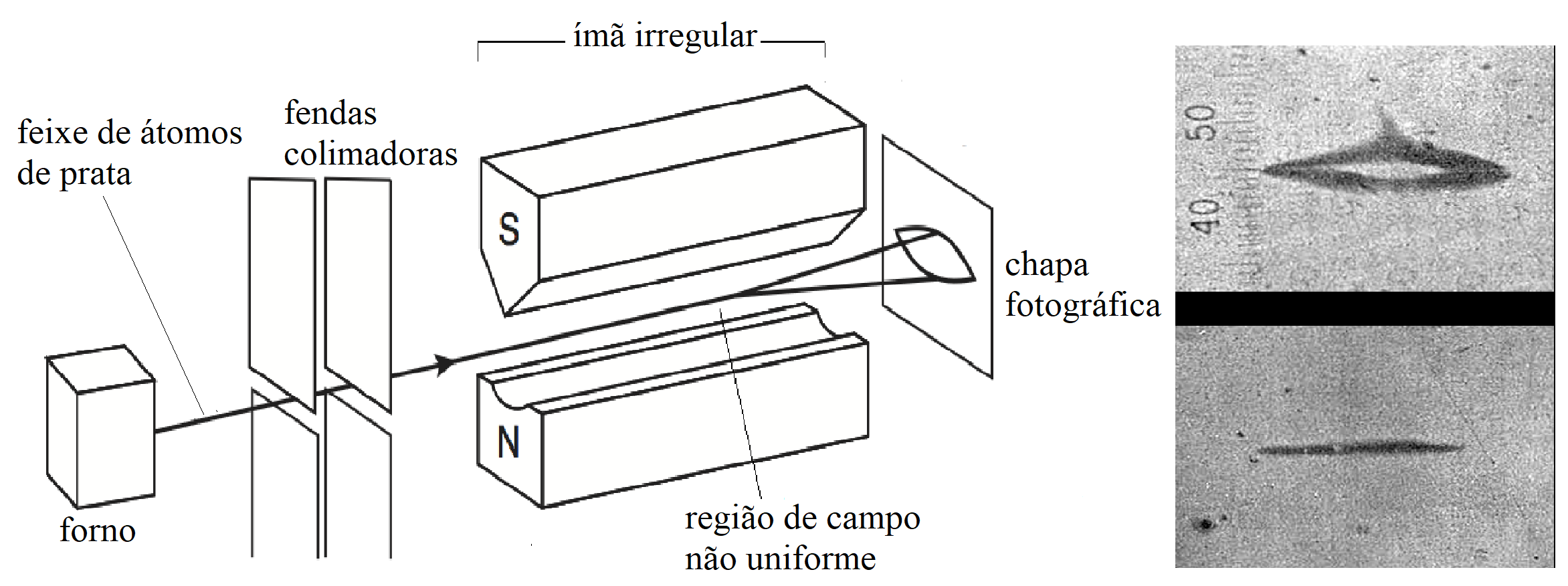}}
\vspace*{-0.1cm}
\caption{\baselineskip=8pt{\small Experimento de Stern-Gerlach.~Na au\-s\^{e}n\-cia (parte inferior) e na presen\c{c}a (parte superior) do campo magn\'{e}tico. Originalmente, a fotografia foi publicada com as linhas dos desdobramentos do feixe inicial na vertical. Na figura, a fotografia foi reproduzida  de acordo com o esquema do experimento (\`{a} esquerda), em que a dire\c{c}\~{a}o do campo magn\'{e}tico \'{e} vertical.}}
\label{stern-gerlach}
\end{figure}

O esquema b\'{a}sico do experimento de Stern-Gerlach consistia em um feixe de \'{a}tomos de pra\-ta direcionado \`{a} uma regi\~{a}o onde havia um cam\-po n\~{a}o uniforme perpendicular \`{a} dire\c{c}\~{a}o do fei\-xe. Um placa fotogr\'{a}fica, colocada a uma certa dis\-t\^{a}n\-cia, permitiria registrar o desdobramento esperado do feixe inicial, devido \`{a} intera\c{c}\~{a}o do momento dipolar dos \'{a}tomos com o campo mag\-n\'{e}\-ti\-co (Figura~\ref{stern-gerlach}).
%

\`{A} \'{e}poca dos experimentos, mesmo com os resultados apresentados mostrando o desdobramento em dois feixes, e a medida do momento magn\'{e}tico da ordem de $10^{-23}$\,J/T,\footnote{\, Esse valor, denotado como $\mu_{_B}$ e, teoricamente, dado por $(e\hbar/2m_e) = 9,\!284\,764\,691\,7\,(29) \times 10^{-24}$\,J/T -- CODATA (2022) --, \'{e} denominado \textit{magneton de Bohr}.}
n\~{a}o se cogitou a hip\'{o}tese do \textit{spin}. Somente ap\'{o}s 1927,\footnote{\, Quando Goudsmit e Uhelenbeck j\'{a} tinha introduzido o conceito de \textit{spin} do el\'{e}tron.}
com o reconhecimento de que os \'{a}tomos de prata no experimento estavam no estado fundamental, para o qual o momento angular orbital e o momento magn\'{e}tico associado s\~{a}o nulos, com\-pre\-en\-deu-se que os resultados dos experimentos deviam-se ao momento magn\'{e}tico in\-tr\'{\i}n\-se\-co do el\'{e}tron, ou seja, ao \textit{spin} do el\'{e}tron. Um ex\-pe\-ri\-men\-to similar, utilizando um feixe de hidrog\^{e}nio no estado fundamental, foi realizado por Thomas Phipps e John B. Taylor~\cite{Phipps}, em 1927, logo ap\'{o}s a introdu\c{c}\~{a}o do conceito de \textit{spin}.

O car\'{a}ter seminal do experimento de Stern-Gerlach, realizado cerca de tr\^{e}s anos antes da introdu\c{c}\~{a}o do conceito de \textit{spin} e da Mec\^{a}nica Qu\^{a}ntica, est\'{a} relacionado ao fato de propiciar a manifesta\c{c}\~{a}o e a verifica\c{c}\~{a}o de v\'{a}rias caracter\'{\i}sticas b\'{a}sicas inesperadas do mundo qu\^{a}ntico -- seja relacionado ao problema da medi\c{c}\~{a}o, ou  ao da prepara\c{c}\~{a}o de estados --,  e ao impacto experimental da t\'{e}cnica de produ\c{c}\~{a}o de feixes moleculares, -- causadora de uma revolu\c{c}\~{a}o nas tecnologias nuclear, at\^{o}mica e molecular.

Antes de Stern e Gerlach, ningu\'{e}m tinha sido capaz de investigar o comportamento dos \'{a}tomos individualmente. Todas essas inova\c{c}\~{o}es estavam presentes no experimento e, assim, gra\c{c}as \`{a} sua simplicidade conceitual, tornou-se o prot\'{o}tipo por excel\^{e}ncia para apresentar e esclarecer diversos pontos da Mec\^{a}nica Qu\^{a}ntica, como o conceito de estado qu\^{a}ntico, da prepara\c{c}\~{a}o de estados, da  interpreta\c{c}\~{a}o  dos processos de me\-di\-\c{c}\~{a}o, e do princ\'{\i}pio da superposi\c{c}\~{a}o de estados.

\vspace*{-0.2cm}
\section{1925 e o \emph{spin}: Goudsmit, Uhlenbeck e o princ\'{\i}pio de exclus\~{a}o de Pauli}

\begin{flushright}
\begin{minipage}{5.5cm}
\baselineskip=10pt {\small
\textit{O Princ\'{\i}pio da Exclus\~{a}o \'{e} estabelecido puramente para o benef\'{\i}cio dos pr\'{o}\-prios el\'{e}trons, que podem ser corrompidos (e se tornar drag\~{o}es ou de\-m\^{o}\-nios) se lhes for permitido se as\-so\-ciar muito livremente.}
\smallskip

\hfill Alan Turing
}
\end{minipage}
\end{flushright}

\vspace*{-0.3cm}
No per\'{\i}odo que se sucedeu \`{a} hip\'{o}tese de Sommerfeld, um novo impulso ocorre na Espectroscopia e na F\'{\i}sica At\^{o}mica. H\'{a} um expressivo incremento na quantidade de pesquisadores atuando nessas \'{a}reas, principalmente  na Europa e nos Estados  Unidos, tanto na parte experimental como na te\'{o}rica, criando um ambiente de troca de informa\c{c}\~{o}es do qual resultou, concomitantemente ao conceito de \textit{spin}, o surgimento da Mec\^{a}nica Qu\^{a}ntica.

Entre 1920 e 1925, o pr\'{o}prio Sommerfeld, Wolfang Pauli e Alfred Land\'{e} se destacam na interpreta\c{c}\~{a}o de vasta quantidade de dados revelados na espectroscopia. Uma grande mul\-tiplicidade de linhas passa a ser revelada nos espectros, na aus\^{e}ncia e na presen\c{c}a de um campo magn\'{e}tico externo. S\~{a}o estabelecidas algumas regras emp\'{\i}ricas que permitiam iden\-tificar as transi\c{c}\~{o}es que poderiam dar origem \`{a}s linhas espectrais -- as chamadas \textit{regras de sele\c{c}\~{a}o}, relativas \`{a}s varia\c{c}\~{o}es permitidas aos n\'{u}meros qu\^{a}n\-ti\-cos. A apresenta\c{c}\~{a}o minuciosa de todas essas regras, determinadas emp\'{\i}rica ou fenomenologicamente, a partir da introdu\c{c}\~{a}o de certos conjuntos de n\'{u}meros inteiros, pode ser encontrada no livro de Sin-itiro Tomonaga~\cite{Tomonaga}, \textit{The Story of Spin}. No entanto, neste ponto, basta lembrar que a manifesta\c{c}\~{a}o do efeito Zeeman an\^{o}malo continuava inexplicada. Nesse contexto, em que os tr\^{e}s n\'{u}meros qu\^{a}nticos, associados a um estado do \'{a}tomo, ainda n\~{a}o eram suficientes para caracterizar as transi\c{c}\~{o}es observadas, um quarto n\'{u}mero qu\^{a}n\-ti\-co, associado a apenas dois valores, \'{e} introduzido \textit{ad hoc}.

Apesar do relativo sucesso para justificar o desdobramento das linhas espectrais sob v\'{a}rias condi\c{c}\~{o}es, que se seguiu \`{a} introdu\c{c}\~{a}o de diversas outras regras emp\'{\i}ricas, ainda sem suporte te\'{o}rico convincente, esse foi um per\'{\i}odo de gran\-de confus\~{a}o. De fato, Sommerfeld, Pauli e Lan\-d\'{e}, al\'{e}m de n\~{a}o utilizarem o mesmo conjunto de n\'{u}meros qu\^{a}nticos, propuseram tamb\'{e}m diferentes modelos para a estrutura at\^{o}mica, como se infere do desabafo de Pauli em resposta a um amigo que comentou que ele n\~{a}o parecia feliz~\cite{Pauli-Science}:
 %
 \vspace*{-0.3cm}
\begin{quotation}
\noindent \baselineskip=10pt {\small\textit{Como se pode estar feliz quando se est\'{a} pensando no efeito Zeeman an\^{o}malo? }}
\end{quotation}

\vspace*{-0.3cm}
Nesse ponto, crucial na compreens\~{a}o do espectro at\^{o}mico e, portanto,  da estrutura at\^{o}mi\-ca, observa-se  que, no caso de um \'{a}tomo n\~{a}o hidrogen\'{o}ide, como o h\'{e}lio, cujo espectro de energia estaria associado a dois pares de n\'{u}meros qu\^{a}nticos, correspondentes aos dois el\'{e}trons, v\'{a}\-rias linhas espectrais que seriam esperadas eram ausentes. Assim, no in\'{\i}cio do ano de 1925, Pau\-li~\cite{Pauli} vislumbra que as linhas suprimidas corresponderiam, principalmente, a estados nos quais os quatro n\'{u}meros qu\^{a}nticos correspondentes a cada el\'{e}tron eram iguais.

Essa observa\c{c}\~{a}o, inicialmente, enunciada co\-mo uma regra para a o\-cor\-r\^{e}n\-cia de um estado at\^{o}\-mi\-co em qualquer \'{a}tomo, independentemente de seu n\'{u}mero a\-t\^{o}\-mi\-co, tornou-se um princ\'{\i}pio fundamental da F\'{\i}sica -- o \textit{prin\-c\'{\i}\-pio de exclus\~{a}o} de Pauli:
 %
\begin{center}
\begin{minipage}{6.5cm}
{\small\textit{Em qualquer \'{a}tomo, n\~{a}o pode haver um estado no qual dois ou mais el\'{e}trons tenham os mesmos quatro n\'{u}meros qu\^{a}nticos.}}
\end{minipage}
\end{center}



\vspace*{-0.3cm}
Imediatamente, a partir desse princ\'{\i}pio, a\-l\'{e}m da explica\c{c}\~{a}o da estrutura at\^{o}mica, Pau\-li encontra a chave que faltava para se compreender, em bases te\'{o}ricas, o enorme suces\-so do arranjo peri\'{o}dico dos elementos qu\'{\i}micos em grupos que t\^{e}m comportamento qu\'{\i}mico semelhantes -- a Tabela Peri\'{o}dica de Mendeleiev.~Se\-gun\-do esse princ\'{\i}pio, a cada n\'{u}mero qu\^{a}ntico principal ($n=1,2,3, \ldots $), associado \`{a} energia, corresponderia uma camada de el\'{e}trons, arranjados em subcamadas com mesma energia, caracterizadas pelo n\'{u}mero qu\^{a}ntico secund\'{a}rio $\big(l = 0,1,2, \ldots, (n-1) \big)$,\footnote{\, Na conven\c{c}\~{a}o atual, ap\'{o}s o surgimento da equa\c{c}\~{a}o de Schr\"{o}dinger.} sendo que para cada subcamada, devido aos dois valores do quarto n\'{u}mero qu\^{a}ntico, o n\'{u}mero de el\'{e}trons seria igual ao dobro do n\'{u}mero qu\^{a}ntico magn\'{e}tico ($m = 0, \pm 1, \pm 2, \ldots , \pm l$), implicando uma satura\c{c}\~{a}o no n\'{u}mero de el\'{e}trons em cada camada, igual a $2\, n^2$. Essa \'{e} a origem da famosa regra de distribui\c{c}\~{a}o eletr\^{o}nica das camadas at\^{o}micas relacionada ao preenchimento das 7 camadas existentes, designadas pelas letras mai\'{u}sculas K, L, M, N, O, P e Q. A mais interna, de menor energia, teria, no m\'{a}ximo 2 el\'{e}trons, e as demais, respectivamente, 8, 18, 32 \textit{etc}.\footnote{\, Esse procedimento, no entanto, n\~{a}o \'{e} suficiente para se definir o preenchimento dos estados at\^{o}micos pelos el\'{e}trons.~O preenchimento correto, com raras ex\-ce\c{c}\~{o}es, \'{e} completado pelas chamadas \textit{regras de Hund}.}

Do ponto de vista experimental, a hip\'{o}tese de Pauli permitia a compreens\~{a}o de todos os resultados obtidos na espectroscopia at\'{e} aquela data, incluindo as manifesta\c{c}\~{o}es da estrutura fina, bem como dos efeitos Zeeman normal e an\^{o}malo. Enquanto os tr\^{e}s primeiros n\'{u}meros qu\^{a}nticos estavam associados \`{a} energia e ao momento angular, e descreviam a estrutura fina e o efeito Zeeman normal, o quarto n\'{u}mero qu\^{a}n\-ti\-co, rec\'{e}m introduzido, e que para Pauli poderia assumir apenas dois valores, apesar de descrever o efeito Zeeman an\^{o}malo, carecia de uma interpreta\c{c}\~{a}o f\'{\i}sica.~Qual grandeza associada ao el\'{e}tron ele representava? Nesse mo\-mento, estava preparado o cen\'{a}rio para a conceitualiza\c{c}\~{a}o de uma nova propriedade dos el\'{e}trons, que seria associada a esse quarto n\'{u}mero qu\^{a}ntico.

Nesse contexto, no mesmo ano de 1925, o jovem Samuel Goudsmit, assistente de Paul Ehrenfest, descobre que o espectro at\^{o}mico poderia ser descrito de modo mais simples e sugestivo, se a esse quarto n\'{u}mero qu\^{a}ntico fossem atribu\'{\i}dos os valores semi-inteiros, iguais a $+1/2$ e $-1/2$.

\`{A} procura de uma interpreta\c{c}\~{a}o f\'{\i}sica para a hip\'{o}tese de Goudsmit, George Uhlenbeck, seu parceiro de pesquisa, sup\~{o}e, ainda de um ponto de vista cl\'{a}ssico, que esse n\'{u}mero estaria associado a um momento angular intr\'{\i}nseco ($\vec S$) do el\'{e}tron, devido a sua rota\c{c}\~{a}o, como se fosse um corpo extenso e, portanto, tendo tamb\'{e}m um momento dipolar intr\'{\i}nseco, que se manifestaria sob a a\c{c}\~{a}o de um campo magn\'{e}tico externo, como ocorria no efeito Zeeman an\^{o}malo e normal~\cite{Goud_Uhle_1, Goud_Uhle_2, Goud, Uhle, Caruso-Oguri_CS}.\footnote{\, Uma parte substancial das hip\'{o}teses de Goudsmit e Uhlenbeck, j\'{a} tinha sido antecipada e, em seguida, descartada por Ralph Kronig, um jovem orientando do pr\'{o}prio Pauli.} A esse mo\-men\-to angular intr\'{\i}nseco, um novo grau de li\-ber\-dade do el\'{e}tron, de maneira simplificada e mnem\^{o}nica, conhecido como \textit{spin}, Goudsmit e Uhlenbeck atribu\'{\i}ram o valor $\hbar/2$ e, ao correspondente momento magn\'{e}tico, o valor $\mu_{_B} = (e \hbar/2m_e)$, igual ao magneton de Bohr. Desse modo, estava finalmente estabelecida a descoberta do \textit{spin}.


De imediato, a hip\'{o}tese  de se associar ao el\'{e}tron uma propriedade intr\'{\i}nseca, baseada, ain\-da, em um modelo mec\^{a}nico, e em argumentos da Mec\^{a}nica e do Eletromagnetismo Cl\'{a}s\-sicos, recebeu v\'{a}rias contesta\c{c}\~{o}es.\footnote{\, Uma das contesta\c{c}\~{o}es, sofrida tamb\'{e}m pelo pr\'{o}prio Kronig, foi a de Lorentz, de que a velocidade ($v$) dos pontos da superf\'{\i}cie de um el\'{e}tron em rota\c{c}\~{a}o com momento angular de \textit{spin} da ordem de $\hbar/2$ seria muito maior que a velocidade da luz no v\'{a}cuo. Com efeito, de acordo com a Mec\^{a}nica Cl\'{a}ssica,
$$ S = \displaystyle \frac{\hbar}{2} = I \omega = \underbrace{(2/5) m_e r_e^2}_{\displaystyle I} \omega $$
implica
$$ v \simeq \frac{\hbar}{m_e r_e} \sim 10^{11}\,{\rm m/s} \gg c $$
em que $I$ \'{e} o momento de in\'{e}rcia, $m_e \simeq 10^{-30}$\,kg e $r_e \simeq 10^{-15}$\,m, respectivamente, a massa e o raio cl\'{a}ssico do el\'{e}tron, e $\omega = v/r_e$, a velocidade angular de rota\c{c}\~{a}o.}
A principal delas era uma suposta contradi\c{c}\~{a}o da  rela\c{c}\~{a}o entre o momento angular e o momento magn\'{e}tico.
 De acordo com o Eletromagnetismo, se a equa\-\c{c}\~{a}o~(\ref{mu}) fosse v\'{a}lida para o \text{spin} ($\hbar/2$)  do el\'{e}tron,
   o momento magn\'{e}tico seria metade do valor proposto por Goudsmit e Uhlenbeck. Ou seja, havia uma discrep\^{a}ncia associada a um fator 2 na express\~{a}o do momento magn\'{e}tico.\footnote{\, Essa aparente contradi\c{c}\~{a}o foi apontada, dentre outros, por Heisenberg, Bohr, Einstein e Pauli.}
As obje\c{c}\~{o}es cessaram quando Llewellyn H. Thomas~\cite{Thomas}, considerando que em rela\c{c}\~{a}o ao el\'{e}tron o pr\'{o}ton em movimento produzia um campo magn\'{e}tico que agia sobre ele, e utilizando a cinem\'{a}tica relativ\'{\i}stica, mostrou que o fator 2 era cancelado na express\~{a}o final do momento magn\'{e}tico. A equa\c{c}\~{a}o~(\ref{mu}), embora v\'{a}lida para o momento angular orbital, n\~{a}o \'{e} v\'{a}lida para o \textit{spin} do el\'{e}tron. A rela\c{c}\~{a}o entre o momento magn\'{e}tico intr\'{\i}nseco ($\vec \mu_e$) e o \textit{spin} ($\vec S$) do el\'{e}tron \'{e} dada por
\begin{equation}\label{Goud}
 \displaystyle \vec \mu_e \, = \, g_e \left( \frac{e}{2m_e}\right) \vec S
\end{equation}
ou,
$$ |\vec \mu_e |\, = \, \frac{|g_e|}{2} \frac{ e \hbar}{2 m_e} \, \simeq \, \mu_{_B}$$
em que $g_e \simeq - 2$ \'{e} o fator $g$ do el\'{e}tron.\footnote{\, O valor do fator $g$ do el\'{e}tron, $g_e = - 2,\!002\,319\,304\,360\,92\,(36)$ -- CODATA (2022), \'{e} uma das mais precisas medidas determinadas na F\'{\i}sica Experimental. Esse fator, associado ao momento magn\'{e}tico de \textit{spin} do el\'{e}tron, muitas vezes \'{e} expresso como um valor positivo $g_s = |g_e|$. Nesse caso, o momento magn\'{e}tico \'{e} expresso como: $ \displaystyle \vec \mu_e \, = \, - g_s \left( \frac{e}{2m_e}\right) \vec S $.}
%
%
%
\section{O surgimento da Mec\^{a}nica Qu\^{a}ntica e a interpreta\c{c}\~{a}o probabil\'{\i}stica de Born}

\begin{flushright}
\begin{minipage}{5.5cm}
\baselineskip=10pt {\small
\textit{Embora nas pesquisas realizadas por Ru\-therford, Bohr, Sommerfeld e outros, a compara\c{c}\~{a}o do \'{a}tomo com um sistema planet\'{a}rio de el\'{e}trons tenha levado a uma interpreta\c{c}\~{a}o qualitativa das propriedades \'{o}pticas e qu\'{\i}micas dos \'{a}tomos, a des\-se\-me\-lhan\c{c}a fundamental entre o espectro at\^{o}\-mico e o espectro cl\'{a}ssico de um sistema de el\'{e}trons im\-p\~{o}e a necessidade de se abandonar o con\-ceito de trajet\'{o}ria para os el\'{e}trons e de se renunciar a uma descri\c{c}\~{a}o visual do \'{a}tomo.
 }
\smallskip

\hfill Werner Heisenberg (1933)

\vspace*{0.5mm}
\hfill  \textit{Nobel Lecture}}

\end{minipage}
\end{flushright}

No mesmo ano de 1925, em que Goudsmit e Uhlenbeck propunham o \textit{spin} do el\'{e}tron, ainda, fortemente, inspirados em um modelo cl\'{a}ssico, Werner Heisenberg~\cite{Heisenberg}, \`{a} \'{e}poca assistente de Max Born, consegue romper com uma das concep\c{c}\~{o}es mais arraigadas no senso comum, e basilar da Mec\^{a}nica Cl\'{a}ssica de Newton, ou seja, que o universo \'{e} um grande mecanismo, regido por leis naturais e matem\'{a}ticas precisas que podem ser utilizadas para prever com total determinismo o comportamento dos corpos celestes e terrestres. Isso se d\'{a} a partir das leis de Newton, em especial da 2\na\ lei da din\^{a}mica, que \'{e} uma equa\c{c}\~{a}o diferencial ordin\'{a}ria de segunda ordem em rela\c{c}\~{a}o ao tempo, cuja solu\-\c{c}\~{a}o para uma part\'{\i}cula, $\big[ x(t), y(t), z(t) \big]$, \'{e} dada pelas equa\c{c}\~{o}es param\'{e}tricas de sua trajet\'{o}ria. Segundo o jovem f\'{\i}sico alem\~{a}o, este conceito dei\-xa de fazer sentido no nova descri\c{c}\~{a}o do microcosmo por ser a trajet\'{o}ria algo inobserv\'{a}vel.


Heisenberg, rejeitando a exist\^{e}ncia do conceito cl\'{a}ssico de trajet\'{o}ria, representa as poss\'{\i}\-veis posi\c{c}\~{o}es e \textit{momenta} da par\-t\'{\i}\-cu\-la  por arranjos num\'{e}ricos cujas opera\c{c}\~{o}es alg\'{e}bricas de multiplica\c{c}\~{a}o n\~{a}o s\~{a}o necessariamente comutativas, como a multiplica\c{c}\~{a}o entre n\'{u}meros reais ou complexos. Ap\'{o}s esses arranjos serem reconhecidos por Born como matrizes, em colabora\c{c}\~{a}o com Pascual Jordan,  estabelecem, propriamente, a primeira vers\~{a}o de uma Mec\^{a}nica Qu\^{a}ntica n\~{a}o relativ\'{\i}stica, ainda sem levar em conta o \textit{spin} do el\'{e}tron~\cite{Born-Heise-Jordan}.

Em rela\c{c}\~{a}o a um sistema cartesiano, as grandezas associadas \`{a} posi\c{c}\~{a}o e ao \textit{momentum},\footnote{\, No sentido lato, entende-se como grandeza toda propriedade f\'{\i}sica \`{a} qual se pode  atribuir uma magnitude, a partir de um processo de medi\c{c}\~{a}o. Por uma influ\^{e}ncia dos precursores da teoria, claramente positivista, as grandezas f\'{\i}sicas utilizadas na Mec\^{a}nica Qu\^{a}ntica s\~{a}o, na maioria das vezes, chamadas de \textit{observ\'{a}veis}.} s\~{a}o representadas por matrizes $x,y,z,p_x, p_y, p_z$, e  o fato de a multiplica\c{c}\~{a}o dessas matrizes n\~{a}o ser comutativa \'{e} expresso pelas chamadas \textit{regras de comuta\c{c}\~{a}o}, dadas por 
$$ \displaystyle  x_i \,  p_j -  p_j \, x_i  = \big[ x_i,  p_j] \, = \, i \hbar\, \delta_{i,j}  \qquad (i, j = 1, 2, 3)
 $$
 em que $\displaystyle \delta_{i,j}\! = \!
\left\{ \!
\begin{array}{l}
1 \ (i=j) \\
0 \ (i\neq j)
\end{array} \!
\right.\!$,
 $\!(x_1\!  =\!   x, x_2\!  =\!  y, x_3\! =\! z)$ e $(p_1 = p_x, p_2 = p_y, p_3 = p_z)$.\footnote{\, Na abordagem matricial, as componentes da po\-si\-\c{c}\~{a}o e do \textit{momentum} n\~{a}o s\~{a}o dadas pelas express\~{o}es cl\'{a}ssicas e n\~{a}o relativ\'{\i}sticas das coordenadas $(x,y,z$) e dos \textit{momenta} $(m v_x, m v_y, m v_z)$, nesse contexto, $x$, $y$, $z$, $p_x$, $p_y$ e $p_z$ s\~{a}o matrizes, ou operadores, definidos pelas regras de comuta\c{c}\~{a}o.}

O momento angular \'{e} introduzido de forma an\'{a}loga \`{a} express\~{a}o cl\'{a}ssica, como o operador matricial cujas componentes s\~{a}o dadas por
 $$\left\{
 \begin{array}{l}
 \displaystyle L_x = y p_z - z p_y  \\
 \displaystyle L_y = z p_x - x p_z   \\
 \displaystyle L_z = x p_y - y p_x
 \end{array}
 \right.
$$
 sujeitas \`{a}s regras de comuta\c{c}\~{a}o
 \renewcommand{\arraystretch}{1.2}
 $$\left\{
 \begin{array}{l}
 \displaystyle \big[L_x, L_y \big] = i \hbar \, L_z \\
 \displaystyle \big[L_y, L_z \big] = i \hbar\, L_x   \\
 \displaystyle \big[L_z, L_x \big] = i \hbar\, L_y
\end{array}
 \right.
 $$
\renewcommand{\arraystretch}{1.}

De maneira tamb\'{e}m similar \`{a} Mec\^{a}nica Cl\'{a}s\-si\-ca, a matriz ($H$) que representa a energia de uma part\'{\i}cula de massa $m$, em um campo conservativo, denominada \textit{matriz hamiltoniana}, \'{e} expressa em fun\c{c}\~{a}o das matrizes associadas \`{a} posi\c{c}\~{a}o e ao  \textit{momentum},
$$ H = \frac{p^2}{2m} \, + \, V(x,y,z) $$
sendo $p^2 = p_x^2 + p_y^2 + p_z^2$, e $V(x,y,z)$ a matriz que representa a energia potencial.

A partir dessas regras de comuta\c{c}\~{a}o fundamentais, Born, Heisenberg, Jordan e Pauli de\-terminam tanto o espectro de energia de uma part\'{\i}cula que oscila harmonicamente com uma dada frequ\^{e}ncia, obtendo a quantiza\c{c}\~{a}o de energia dos osciladores de Planck, como o espectro de energia do hidrog\^{e}nio, com sua estrutura fina, e sob a\c{c}\~{a}o de campos el\'{e}tricos -- o efeito Stark --, e magn\'{e}ticos -- o efeito Zeeman normal.

Do ponto de vista matem\'{a}tico, os resultados s\~{a}o obtidos diagonalizando a matriz hamiltoniana, sendo seus autovalores identificados por Born, como os poss\'{\i}veis valores de energia mas, como aponta o matem\'{a}tico Bartel L.~van der Waerden~\cite{Waerden}:

\vspace*{-0.3cm}
\begin{quotation}
\noindent \baselineskip=10pt {\small \textit{
No entanto, ele n\~{a}o percebeu que os autovetores determinam os estados estacion\'{a}rios do \'{a}tomo; ele os usou apenas como aux\'{\i}lio matem\'{a}\-ti\-co para realizar a transforma\c{c}\~{a}o para os eixos principais.~O significado f\'{\i}\-si\-co dos autovetores n\~{a}o foi esclare\-ci\-do antes de Schr\"{o}\-din\-ger.}}
\end{quotation}

\vspace*{-0.3cm}
Born, Heisenberg e Jordan desenvolvem a teoria de modo a incluir a din\^{a}mica dos el\'{e}trons at\^{o}micos perturbados por campos eletro\-mag\-n\'{e}\-ti\-cos dependentes do tempo, abordando tamb\'{e}m os \'{a}tomos com muitos el\'{e}trons, mas sem estend\^{e}-la aos processos aperi\'{o}dicos.


A segunda vers\~{a}o, obtida por Erwin Schr\"{o}\-dinger, foi motivada, principalmente, pela hip\'{o}\-te\-se de Louis de Broglie, apresentada como tese de seu doutorado, em 1924~\cite{de_Broglie}. Como Einstein havia mostrado, em sua interpreta\c{c}\~{a}o do problema da Radia\c{c}\~{a}o de Corpo Negro, era poss\'{\i}vel associar a uma onda eletromagn\'{e}tica plana monocrom\'{a}tica,  um conjunto de part\'{\i}culas que portavam, cada uma, um fragmento ou \textit{quantum} de energia $E$, proporcional \`{a} frequ\^{e}ncia ($\nu$) da radia\c{c}\~{a}o, dada pela express\~{a}o de Planck; $E = h \nu$. Nesse sentido, uma onda eletromagn\'{e}tica apresentaria uma natureza discreta, sendo constitu\'{\i}da de corp\'{u}sculos n\~{a}o materiais de energia: os f\'{o}tons. De maneira an\'{a}loga, de Broglie considera que, assim como a um conjunto de f\'{o}tons associa-se uma onda eletromagn\'{e}tica, pode-se associar a um feixe homog\^{e}neo de part\'{\i}culas materiais, cada qual com \textit{momentum} $p$, um comportamento ondulat\'{o}rio, caracterizado por uma onda progressiva de comprimento de onda $\lambda = h/p$.\footnote{\, Para de Broglie, o \textit{momentum} da part\'{\i}cula seria dado pela express\~{a}o relativ\'{\i}stica, $p = \displaystyle \frac{m v}{\sqrt{1 - v^2/c^2}}$.}

Schr\"{o}dinger, ent\~{a}o, em 1926, publica uma s\'{e}rie de artigos nos quais estabelece a vers\~{a}o ondulat\'{o}ria da Mec\^{a}nica Qu\^{a}ntica~\cite{Sch_I_IV}, mostrando tamb\'{e}m  a equival\^{e}ncia entre as vers\~{o}es matricial e ondulat\'{o}ria.
Inicialmente, ao abordar o problema do \'{a}tomo, considera que, em vez de uma onda progressiva, as vibra\c{c}\~{o}es at\^{o}micas, por serem espacialmente limitadas, estariam associadas a ondas estacion\'{a}rias. Desse modo, a quantiza\c{c}\~{a}o da energia resultava de um problema de autovalor, envolvendo uma equa\c{c}\~{a}o diferencial parcial e um campo escalar, $\psi (x,y,z)$, sujeito a condi\c{c}\~{o}es de contorno apropriadas, semelhante ao problema da corda vibrante. Atualmente, a vers\~{a}o ondulat\'{o}ria \'{e} apresentada de maneira an\'{a}loga \`{a} formula\c{c}\~{a}o matricial; o operador hamiltoniano ($H$) que representa a energia de uma part\'{\i}cula de massa $m$, em um campo conservativo, \'{e} expresso como na Mec\^{a}nica Cl\'{a}ssica, em termos dos operadores associados \`{a} posi\c{c}\~{a}o e ao \textit{momentum}. Enquanto os operadores de posi\c{c}\~{a}o s\~{a}o, simplesmente, as coordenadas cartesianas $(x,y,z)$, os associados \`{a}s componentes do \textit{momentum} s\~{a}o operadores diferenciais dados por
$$\displaystyle  p_x = - i \, \hbar \, \frac{\partial}{\partial x}, \quad  p_y = - i \, \hbar \, \frac{\partial}{\partial y},  \quad p_z = - i \, \hbar \, \frac{\partial}{\partial z} $$
e o espectro de energia -- o conjunto de poss\'{\i}veis energias $\{E\}$ --, determinado pela equa\c{c}\~{a}o diferencial de autovalor
$$\displaystyle  H\, \psi =  \left[ \frac{p^2}{2m} \, + \, V(x,y,z) \right] \psi(x, y, z) = E \, \psi $$

Apesar da equa\c{c}\~{a}o de autovalor proporcionar a solu\c{c}\~{a}o do movimento limitado de uma part\'{\i}cula em um campo conservativo, como o el\'{e}tron no \'{a}tomo ou o oscilador harm\^{o}nico, o f\'{\i}sico austr\'{\i}aco obt\'{e}m tamb\'{e}m a equa\c{c}\~{a}o que descreve a evolu\c{c}\~{a}o espa\c{c}o-temporal da part\'{\i}\-cu\-la, tanto no interior do \'{a}tomo, como em movimento n\~{a}o limitado sob a a\c{c}\~{a}o de campos n\~{a}o conservativos dependentes do tempo ($t$) --  a \textit{equa\-\c{c}\~{a}o de onda de Schr\"{o}dinger} ou, simplesmente, \textit{equa\c{c}\~{a}o de Schr\"{o}dinger},
$$ \displaystyle i\,\hbar\, \frac{\partial }{\partial t} \Psi(x,y,z,t) \, = \, \left[ \frac{p^2}{2m} \, + \, V(x,y,z,t) \right]
\Psi $$
em que $V(x,y,z,t)$ \'{e} a energia potencial, e $\Psi$, devido \`{a} analogia de L.~de Broglie, \'{e} chamada \textit{fun\c{c}\~{a}o de onda} de Schr\"{o}dinger.

A formula\c{c}\~{a}o ondulat\'{o}ria \'{e} estendida tam\-b\'{e}m aos \'{a}tomos multi-eletr\^{o}nicos, mas as duas ver\-s\~{o}es ainda careciam de um cunho interpretativo. Tan\-to as autofun\c{c}\~{o}es do operador hamiltoniano, co\-mo a fun\c{c}\~{a}o de onda eram apenas campos escalares auxiliares, n\~{a}o representando qualquer propriedade ou grandeza f\'{\i}sica. Apesar de ter tentado conceber uma interpreta\c{c}\~{a}o f\'{\i}sica para a fun\c{c}\~{a}o de onda, Schr\"{o}dinger n\~{a}o obt\'{e}m sucesso. Al\'{e}m da extensa similaridade formal com a Mec\^{a}nica Cl\'{a}ssica, em particular com a formula\c{c}\~{a}o ana\-l\'{\i}\-ti\-ca de Hamilton, o car\'{a}ter determin\'{\i}stico im\-pl\'{\i}\-ci\-to na Mec\^{a}nica, e mesmo na Re\-la\-ti\-vidade Restrita, ainda predominava ao se buscar uma interpreta\c{c}\~{a}o da teoria.~A ruptura com o determinismo cl\'{a}ssi\-co de Laplace s\'{o} ocorre com a interpreta\c{c}\~{a}o probabil\'{\i}stica proposta por Max Born.

Ao analisar o espalhamento de um feixe homog\^{e}neo de el\'{e}trons por um \'{a}tomo, Born~\cite{Born_espalha}, inicialmente, antes de qualquer intera\c{c}\~{a}o, representa as part\'{\i}culas incidentes pela autofun\c{c}\~{a}o ($\psi_i = \phi_{_E}$) associada a uma part\'{\i}cula livre com uma dada energia $E$. Ap\'{o}s a intera\c{c}\~{a}o, a solu\c{c}\~{a}o ($\psi_{_f}$) da equa\c{c}\~{a}o de Schr\"{o}dinger assintoticamen\-te, no infinito, pode ser expressa por uma superposi\c{c}\~{a}o das autofun\c{c}\~{o}es correspondentes ao espectro de energia da part\'{\i}cula livre,
$$ \psi_{_f} (x,y,z) = \int c(E') \, \phi_{_{E'}} (x,y,z) \, {\rm d}x \, {\rm d}x \, {\rm d}z $$
cujos coeficientes determinam as probabilidades de um el\'{e}tron inicialmente com energia $E$, em um estado $\phi_{_E}$, ser encontrado em um estado $\phi_{_{E'}}$ com energia $E'$. Os coeficientes s\~{a}o dados por\footnote{\, Na linguagem da \'{A}lgebra Linear, pelo produto escalar $\big( \phi_{_{E'}}, \psi_{_f} \big)$.}
$$\displaystyle c(E') = \int \phi_{_{E'}}^\ast \, \psi_{_f} \, {\rm d}x \, {\rm d}x \, {\rm d}z  = \big( \phi_{_{E'}}, \psi_{_f} \big) $$
 e, as probabilidades, segundo Born, por\footnote{\, Se as autofun\c{c}\~{o}es, ou autoestados de energia, s\~{a}o devidamente normalizadas como
$$ \displaystyle  \int \phi_{_{E'}}^\ast  \, \phi_{_E}  \, {\rm d}x \, {\rm d}x \, {\rm d}z = \big( \phi_{_{E'}}, \phi_{_E} \big) = \delta (E - E') $$
sendo $\delta (E-E')$ a fun\c{c}\~{a}o delta de Dirac, definida pe\-la rela\c{c}\~{a}o
$$ \displaystyle \int f(E) \, \delta (E-E') \,  {\rm d}E = f(E')$$
 }
$$ P (E') = \big| c(E') \big|^2 = \left|\big( \phi_{_{E'}}, \psi_{_f} \big) \right|^2$$

Uma das principais consequ\^{e}ncias  do prin\-c\'{\i}\-pio de Pauli, ap\'{o}s o surgimento da formula\c{c}\~{a}o de Schr\"{o}dinger, foi a generaliza\c{c}\~{a}o feita por Paul Dirac, em 1926, acerca do comportamento de sistemas de part\'{\i}culas de mesma esp\'{e}cie, em particular de gases ideais.
Segundo Dirac~\cite{Dirac_stat}, a fun\c{c}\~{a}o de onda de um par de part\'{\i}culas (1,2), $\Psi(1,2)$, pode ser expressa como uma combina\-\c{c}\~{a}o sim\'{e}trica, das autofun\c{c}\~{o}es que representam seus estados estacion\'{a}rios, pela troca de part\'{\i}\-cu\-las,
$$  \Psi(1,2) \ \propto \ \psi_\alpha (1) \, \psi_\beta (2) \,+ \,\psi_\beta (1) \psi_\alpha (2)$$
ou, por uma combina\c{c}\~{a}o antissim\'{e}trica,
\begin{eqnarray*}
  \Psi(1,2) &\propto& \psi_\alpha (1) \, \psi_\beta (2)\, -\, \psi_\beta (1) \psi_\alpha (2)  \\
   &\propto& {\rm det} \left(
                                                                                                                   \begin{array}{cc}
                                                                                                                    \psi_\alpha (1)  & \psi_\alpha (2) \\
                                                                                                                     \psi_\beta (1)  & \psi_\beta (2)  \\
                                                                                                                   \end{array}
                                                                                                                 \right)
\end{eqnarray*}
em que $\alpha$ e $\beta$ s\~{a}o dois conjuntos de n\'{u}meros qu\^{a}nticos associados \`{a}s autofun\c{c}\~{o}es.

Para um n\'{u}mero $N$ de part\'{\i}culas, a combina\c{c}\~{a}o antissim\'{e}trica pode ser escrita  como determinante da matriz ($N \times N$),
{\small $$ \Psi(1,2  \ldots N)\! \propto\! {\rm det}\! \left(\! \!
                                                      \begin{array}{cccc}
                                                           \psi_\alpha (1)  & \psi_\alpha (2) & \ldots & \psi_\alpha (N)\\
                                                           \psi_\beta (1)  & \psi_\beta (2) & \ldots & \psi_\beta (N) \\
                                                           \vdots   &  \vdots & \vdots  &\vdots \\
                                                           \psi_\zeta (1)  & \psi_\zeta (2) & \ldots & \psi_\zeta (N) \\
                                                      \end{array} \! \!
                                                       \right)
$$}

\noindent a qual se anula quando duas ou mais part\'{\i}culas s\~{a}o associadas ao mesmo conjunto de n\'{u}meros qu\^{a}nticos.

Assim, Dirac estabelece que a fun\c{c}\~{a}o de onda de um sistema de part\'{\i}culas que obedecem ao princ\'{\i}pio de Pauli, como os el\'{e}trons, deve ser antissim\'{e}trica. Aplicando essa propriedade a um g\'{a}s de part\'{\i}culas que obedecem ao princ\'{\i}pio de Pauli, Dirac  obt\'{e}m a distribui\c{c}\~{a}o das part\'{\i}culas em fun\c{c}\~{a}o da energia, a qual na mesma \'{e}poca foi obtida por Enrico Fermi~\cite{Fermi_stat}, baseando-se no fato de que as part\'{\i}culas de um g\'{a}s n\~{a}o relativ\'{\i}stico, de acordo com o  princ\'{\i}pio de Pauli, n\~{a}o podiam compartilhar estados de mesma energia, -- a \textit{distribui\c{c}\~{a}o de Fermi-Dirac}.

A solu\c{c}\~{a}o sim\'{e}trica, que n\~{a}o impunha restri\c{c}\~{o}es ao n\'{u}mero de part\'{\i}culas em cada estado, representaria um g\'{a}s de part\'{\i}culas
sem massa como os f\'{o}tons, obedecendo  \`{a} distribui\c{c}\~{a}o de Planck, ou massivas n\~{a}o relativ\'{\i}sticas, que obedecessem \`{a} distribui\c{c}\~{a}o obtida por Einstein~\cite{Einstein_stat} -- a \textit{distribui\c{c}\~{a}o de Bose-Einstein}, a partir da abordagem de Satyandranath Bose~\cite{Bose}.\footnote{\, Einstein obteve a chamada distribui\c{c}\~{a}o de Bose-Einstein ao estender a abordagem de Bose, originalmente aplicada a um g\'{a}s de f\'{o}tons, aos gases de part\'{\i}culas massivas e n\~{a}o relativ\'{\i}sticas com \textit{spins} n\~{a}o fracion\'{a}rios, apesar de, na \'{e}poca,  n\~{a}o  haver o conceito de \textit{spin}.}

\section{As matrizes de Pauli e a re\-presenta\c{c}\~{a}o do \emph{spin} }\label{Matrizes_Pauli}

\begin{flushright}
\begin{minipage}{5.5cm}
\baselineskip=10pt {\small
\textit{Em todos os aspectos, \'{e} realmente estranho que absolutamente ningu\'{e}m tenha proposto, at\'{e} o trabalho de Pauli {\rm{[...]}} e Dirac, que \'{e} vinte anos depois da Relatividade Especial {\rm{[...]}}, este relato sinistro de que uma tribo misteriosa chamada fam\'{\i}lia dos espinores habita o espa\c{c}o isotr\'{o}pico {\rm{[}}tridimensional{\rm{]}} ou o mundo de Einstein-Minkows\-ki {\rm [}quadridimensional{\rm ]}.
 }
\smallskip

\hfill Sin-Itiro Tomonaga}

\end{minipage}
\end{flushright}

Apesar do ceticismo e das cr\'{\i}ticas iniciais, Wolfang Pauli foi uma figura central no estabelecimento e na inclus\~{a}o do {\textit{spin} na Mec\^{a}nica Qu\^{a}ntica n\~{a}o relativ\'{\i}stica. Com o intuito de incorporar a hip\'{o}tese, de Goudsmit e Uhlenbeck, de momento angular intr\'{\i}nseco semi-inteiro do el\'{e}tron, na formula\c{c}\~{a}o de Schr\"{o}dinger,  Pauli ~\cite{Pauli_spin}, em 1927, em vez de um campo escalar, considera a fun\c{c}\~{a}o de onda um  campo ($\psi$) representado por uma matriz coluna $(2 \times 1)$,
$$ \psi = \left( \! \begin{array}{c}
       \psi^+ \\
       \psi^-
\end{array}\! \right)
$$
e as componentes cartesianas do operador de \textit{spin}, matrizes quadradas $(2 \times 2)$, que satisfazem as mesmas rela\c{c}\~{o}es de comuta\c{c}\~{a}o obedecidas pelo  momento angular orbital, ou seja,
 \renewcommand{\arraystretch}{1.2}
$$\left\{
 \begin{array}{l}
 \displaystyle \big[S_x, S_y \big] = i \hbar \, S_z \\
 \displaystyle \big[S_y, S_z \big] = i \hbar\, S_x  \\
 \displaystyle \big[S_z, S_x \big] = i \hbar\, S_y
 \end{array}
 \right.
 $$
\renewcommand{\arraystretch}{1.}

Operacionalmente, uma representa\c{c}\~{a}o ex\-pl\'{\i}\-ci\-ta para o \textit{spin} $1/2$ do el\'{e}tron \'{e} definida em termos das  chamadas  \textit{matrizes de Pauli},\footnote{\, Os autovalores de cada matriz de Pauli s\~{a}o $+1$ e $-1$, e os respectivos autovetores normalizados dados por
\begin{center}
\renewcommand{\arraystretch}{1.}
\begin{tabular}{ l | l  | l   }
autovalores & \quad \qquad $+1$ & \quad \qquad $- 1$ \\
\hline
                    & $\displaystyle  \chi_{_x}^+ = \frac{1}{\sqrt 2} \left( \! \begin{array}{c}
                                                                                 1 \\
                                                                                 1
                                                                           \end{array} \! \right) $
                    & $\displaystyle  \chi_{_x}^- = \frac{1}{\sqrt 2} \left( \! \begin{array}{c}
                                                                              \!  \! {\mbox{-}} 1 \\                                                                                 1
                                                                           \end{array} \! \right) $ \\ [0.3cm]
\cline{2 - 3}  
autovetores & $\displaystyle  \chi_{_y}^+ = \frac{1}{\sqrt 2} \left( \! \begin{array}{c}
                                                                                 1 \\
                                                                                 i
                                                                           \end{array} \! \right) $
                    & $\displaystyle  \chi_{_y}^-= \frac{1}{\sqrt 2} \left( \! \begin{array}{c}
                                                                                 i \\
                                                                                 1
                                                                           \end{array}\! \right) $ \\ [0.3cm]
\cline{2 - 3} 
                   & $\displaystyle  \chi_{_z}^+ =  \left( \! \begin{array}{c}
                                                                                 1 \\
                                                                                 0
                                                                           \end{array} \! \right) $
                    & $\displaystyle  \chi_{_z}^- = \ \left(\! \begin{array}{c}
                                                                                 0 \\
                                                                                 1
                                                                           \end{array} \! \right)  $
\end{tabular}
\end{center}
\renewcommand{\arraystretch}{1.}
}
\renewcommand{\arraystretch}{1.3}
$$ \left\{
\begin{array}{l}
\sigma_x \! =\! \left( \begin{array}{cc}
                                     0   & 1 \\
                                     1   &  0
                            \end{array} \right) \\
       \ \\                      
\sigma_y \!=\! \left(  \begin{array}{cc}
                                     0  & - i \\
                                     i   &  0
                            \end{array} \right) \\
       \  \\                   
\sigma_z \! = \!\left(  \begin{array}{cc}
                                    1  & 0 \\
                                    0  &  -1
                             \end{array} \right)
\end{array}
\right.
$$
\renewcommand{\arraystretch}{1.}
em que as componentes s\~{a}o dadas por
$$
S_x = \frac{\hbar}{2} \, \sigma_x, \qquad  S_y = \frac{\hbar}{2} \, \sigma_y, \qquad  S_z = \frac{\hbar}{2} \, \sigma_z
$$
e
$$ \vec S =  \big( S_x, S_y, S_z\big) = \frac{\hbar}{2} \, \big( \sigma_x, \sigma_y, \sigma_z\big) = \frac{\hbar}{2} \, \vec \sigma  $$

Apesar da representa\c{c}\~{a}o matricial, a fun\c{c}\~{a}o de onda, segundo Pauli, n\~{a}o \'{e} um campo vetorial, \'{e} um \textit{campo espinorial} de ordem 1, geralmente chamado \textit{espinor de Pauli}, cujas componentes, em uma mudan\c{c}a de sistemas cartesianos, correspondente a uma rota\c{c}\~{a}o de um \^{a}ngulo $\theta$ em torno do eixo $z$, se transformam como\footnote{\, Inicialmente, os eixos dos sistemas de eixos ($x,y,z$) e ($x',y',z'$) s\~{a}o paralelos.}
$$ \displaystyle \psi' = e^{i S_z \theta/\hbar}\, \psi =   \left(\!  \begin{array}{cc}
                                    e^{i \theta/2}  & 0 \\
                                    0  &  e^{- i \theta/2}
                             \end{array}\! \right)
\left( \! \begin{array}{c}
       \psi^+ \\
       \psi^-
\end{array}\! \right)
$$
em que as componentes do \textit{spin}, ou as correspondentes matrizes de Pauli, s\~{a}o os chamados geradores da representa\c{c}\~{a}o espinorial bi-dimensional do grupo das rota\c{c}\~{o}es.\footnote{\, Rota\c{c}\~{o}es em torno dos eixos $x$ e $y$, correspondentes aos \^{a}ngulos $\alpha$ e $\beta$ s\~{a}o expressas como
$$\scriptsize   \left(\! \! \begin{array}{cc}
                                    \cos (\alpha/2)   &  i\, {\rm sen}\, (\alpha/2) \\
                                  i\, {\rm sen}\, (\alpha/2)  &  \cos (\alpha/2)
                             \end{array}\! \!\right) \ \mbox{e} \
\left(\! \! \begin{array}{cc}
                                    \cos (\beta/2)   &   {\rm sen}\, (\beta/2) \\
                                   {\rm - sen}\, (\beta/2)  & \cos  (\beta/2)
                             \end{array}\! \!\right)
$$
   }

\subsection{A precess\~{a}o de Larmor segun\-do Pauli}

\begin{flushright}
\begin{minipage}{5.5cm}
\baselineskip=10pt {\small
\textit{Na mec\^{a}nica qu\^{a}ntica, n\~{a}o se pode falar sobre a {\rm dire\c{c}\~{a}o} do momento angular no mesmo sentido que se faz classicamente; no entanto, h\'{a} uma analogia muito pr\'{o}xima -- t\~{a}o pr\'{o}xima que continuamos a cham\'{a}-la de precess\~{a}o.
 }
\smallskip

\hfill Richard Feynman}

\end{minipage}
\end{flushright}


Uma aplica\c{c}\~{a}o t\'{\i}pica do formalismo de Pauli \'{e} a an\'{a}lise do comportamento do el\'{e}tron sob a a\c{c}\~{a}o de um campo magn\'{e}tico uniforme. De acordo com Pauli, o momento magn\'{e}tico de \textit{spin} ($\vec \mu_e$) do el\'{e}tron pode ser  expresso como
$$\displaystyle \vec \mu_e\! = \! g_e \left( \frac{e}{2 m_e} \right) \, \vec S\! =\! - \frac{|g_e|}{2} \underbrace{\bigg( \frac{e \hbar}{2m_e} \bigg)}_{\mu_{_B}}   \vec \sigma \! =\!   - \gamma_e \, \frac{\hbar}{2} \vec \sigma $$
em que  $\displaystyle \gamma_e = |g_e|\, \frac{e}{2 m_e} \simeq 1,\!76 \times 10^{11}$\,s$^{-1}\cdot$T$^{-1}$  \'{e} a \textit{raz\~{a}o giromagn\'{e}tica} do el\'{e}tron.\footnote{\, $\displaystyle \gamma_e = 1,\!760\,859\,627\,84\,(55) \times 10^{11}$\,s$^{-1}\cdot$T$^{-1}$ -- CODATA (2022).}

Supondo que o el\'{e}tron \'{e} oriundo de um dos feixes que emergem de um aparato de Stern-Gerlach e o campo ($\vec B$) \'{e} aplicado em uma di\-re\-\c{c}\~{a}o $z$, a hamiltoniana ($H$) do el\'{e}tron,  levando em conta apenas o grau de liberdade de \textit{spin}, \'{e} dada por
$$  H = - \vec \mu_e \cdot \vec B = \big( \gamma_e  B \big)  \frac{\hbar}{2} \, \sigma_z = \omega_{_L} \frac{\hbar}{2} \left( \begin{array}{cc}
       1 & 0 \\
       0 &  -1
\end{array} \right)
$$
em que $\omega_{_L} = \gamma_e B$ \'{e} a frequ\^{e}ncia de Larmor.

Nesse caso, a equa\c{c}\~{a}o de Schr\"{o}dinger \'{e} expressa como
\begin{eqnarray*}
   i \hbar \frac{\rm d}{{\rm d}t} \left( \! \begin{array}{c}
       \psi^+ \\
       \psi^-
\end{array}\! \right) &=& H  \left( \! \begin{array}{c}
       \psi^+ \\
       \psi^-
\end{array}\! \right)  \\
   &=&  \omega_{_L} \frac{\hbar}{2} \left(  \begin{array}{cc}
       1  & 0 \\
       0  &  -1
\end{array} \right)  \left( \! \begin{array}{c}
       \psi^+ \\
       \psi^-
\end{array} \! \right)
\end{eqnarray*}
ou, equivalentemente, por um sistema de equa\c{c}\~{o}es independentes,
$$
\left\{ \begin{array}{l}
\displaystyle i \frac{\rm d}{{\rm d} t}  \psi^+ =  \frac{\omega_{_L}}{2} \psi^+ \\
\ \\
\displaystyle i \frac{\rm d}{{\rm d} t} \psi^- = \frac{\omega_{_L}}{2} \psi^-
\end{array} \right. \ \  \Rightarrow \ \
\left\{ \begin{array}{l}
 \displaystyle \psi^+ =  C^+ e^{- i \,  \omega_{_L} t/2}  \\
 \ \\
 \displaystyle\psi^- =  C^- e^{ i \,  \omega_{_L} t/2}
\end{array} \right.
$$
cuja solu\c{c}\~{a}o geral \'{e}
$$
\psi (t) = \left( \! \begin{array}{l}
 \displaystyle C^+ e^{- i \,  \omega_{_L} t/2}  \\
 \displaystyle C^- \, e^{ i \,  \omega_{_L} t/2}
\end{array}\! \right)
$$

Considerando que o campo magn\'{e}tico inter\-no no aparato de Stern-Gerlach \'{e} ortogonal ao cam\-po magn\'{e}tico uniforme, na dire\c{c}\~{a}o $x$, e o feixe selecionado corresponde ao autovalor positivo do \textit{spin}, o estado inicial do el\'{e}tron \'{e} o autovetor $\chi^+_x$ da componente $S_x$ , ou seja,
$$
\psi(0) = \displaystyle \chi^+_x = \frac{1}{\sqrt 2} \left( \! \begin{array}{c}
                                                                                 1 \\
                                                                                 1
                                                                           \end{array} \! \right)
\quad \Rightarrow \quad  C^+ =  C^- = \frac{1}{\sqrt 2}
$$
implica
$$ \psi (t) = \displaystyle   \frac{1}{\sqrt 2} \left( \! \begin{array}{l}
   e^{- i \,  \omega_{_L} t/2}  \\
 \, e^{ i  \, \omega_{_L} t/2}
\end{array}\! \right)
$$
Assim, o estado final do el\'{e}tron n\~{a}o est\'{a} associado a uma componente espec\'{\i}fica do \textit{spin}, sendo uma superposi\c{c}\~{a}o dos
autovetores de qualquer das componentes $S_x$, $S_y$ e $S_z$, e as pro\-babilidades associadas aos respectivos autovalores s\~{a}o, respectivamente,\footnote{\, Nesse caso, o produto escalar \'{e} dado pelo produto $(\chi^\dagger \psi)$. }
{\small
$$ \left\{ \!
\begin{array}{l}
\displaystyle
 P(\hbar/2)_x = \left| {\chi^+_x}^\dagger \psi(t) \right|^2  \\
  \displaystyle \qquad = \left| \frac{1}{2} \big( 1 \quad 1 \big) \left( \! \begin{array}{l}
   e^{- i \,  \omega_{_L} t/2}  \\
 \, e^{ i  \, \omega_{_L} t/2}
\end{array}\! \right) \right|^2 \\
  \displaystyle \qquad = \left| \frac{e^{- i \,  \omega_{_L} t/2} + e^{ i \,  \omega_{_L} t/2}}{2} \right|^2 = \cos^2  \frac{\omega_{_L} t}{2} \\
\ \\
\displaystyle
 P(-\hbar/2)_x = 1 -  P(\hbar/2)_x = \, {\rm sen}^2 \, \frac{\omega_{_L} t}{2}
 \end{array}
 \right.
 $$}
$$ \left\{ \!
\begin{array}{l}
\displaystyle
 P(\hbar/2)_y = \left| {\chi^+_y}^\dagger \psi(t) \right|^2 \\
 \displaystyle \quad = \!\left| \frac{1}{2} \big( 1 \ \ \mbox{-} i  \big) \left( \! \begin{array}{l}
   e^{- i \,  \omega_{_L} t/2}  \\
 \, e^{ i  \, \omega_{_L} t/2}
\end{array}\! \right) \right|^2 \\
 \displaystyle \quad = \! \left| \frac{e^{- i \,  \omega_{_L} t} - i\, e^{ i \,  \omega_{_L} t/2}}{2} \right|^2 \! = \frac{1 + \, {\rm sen}\,  2  \omega_{_L} t/2}{2} \\
\ \\
\displaystyle
 P(-\hbar/2)_y = 1 -  P(\hbar/2)_y = \frac{1 - \, {\rm sen}\,  2  \omega_{_L} t/2}{2}
  \end{array}
 \right.
 $$
$$ \left\{\!
\begin{array}{l}
\displaystyle
 P(\hbar/2)_z = \left| {\chi^+_z}^\dagger \psi(t) \right|^2 \\
 \displaystyle \qquad = \left| \frac{1}{\sqrt 2} \big( 1 \quad  0 \big) \left( \! \begin{array}{l}
   e^{- i \,  \omega_{_L} t}  \\
 \, e^{ i  \, \omega_{_L} t}
\end{array}\! \right) \right|^2 \\
\displaystyle \qquad = \frac{1}{2} \left| e^{- i \,  \omega_{_L} t/2} \right|^2 = \frac{1}{2} \\
\ \\
\displaystyle
 P(-\hbar/2)_z = 1 -  P(\hbar/2)_z = \frac{1}{2}
 \end{array}
 \right.
 $$

 Desse modo, os respectivos valores m\'{e}dios de $S_x$, $S_y$ e $S_z$ variam no tempo como
$$ \left\{ \!
\begin{array}{l}
\displaystyle
\langle S_x \rangle =  P(\hbar/2)_x \left( \frac{\hbar}{2} \right) +  P(- \hbar/2)_x \left(- \frac{\hbar}{2} \right) \\
\displaystyle \qquad = \frac{\hbar}{2} \left( \cos^2  \frac{\omega_{_L} t}{2} - \, {\rm sen}^2\,  \frac{ \omega_{_L} t}{2} \right) \\
\displaystyle \qquad  = \frac{\hbar}{2} \cos \omega_{_L} t \\
\ \\
\displaystyle
\langle S_y \rangle =  P(\hbar/2)_y \left( \frac{\hbar}{2} \right) +  P(- \hbar/2)_y \left(- \frac{\hbar}{2} \right) \\
\displaystyle \qquad = \frac{\hbar}{2} \left( \frac{1 +  \, {\rm sen}\,  \omega_{_L} t}{2} -  \frac{1 -  \, {\rm sen}\,  \omega_{_L} t}{2} \right) \\
\displaystyle \qquad  = \frac{\hbar}{2}  \, {\rm sen}\,  \omega_{_L} t \\
\ \\
\displaystyle
\langle S_z \rangle \!=\!  P(\hbar/2)_z \left( \frac{\hbar}{2} \right) +  P(- \hbar/2)_z \left(- \frac{\hbar}{2} \right)\! =\! 0
 \end{array}
 \right.
 $$

De maneira similar ao caso do momento mag\-n\'{e}\-ti\-co cl\'{a}ssico (Ap\^{e}ndice \ref{Larmor_classica}), se o estado inicial n\~{a}o for um autovetor da componente do \textit{spin} na dire\c{c}\~{a}o do campo, os valores m\'{e}dios do momento magn\'{e}tico de \textit{spin} oscilam com a frequ\^{e}ncia de Larmor nas dire\c{c}\~{o}es  perpendiculares ao campo, enquanto na dire\c{c}\~{a}o do campo ou \'{e} nulo ou se mant\'{e}m constante.

De acordo com o valor da raz\~{a}o giromag\-n\'{e}\-ti\-ca do el\'{e}tron $\big(\gamma_e \simeq 10^{11}$\,s$^{-1}\cdot$T$^{-1}\big)$,
a frequ\^{e}ncia de Larmor ($\nu_{_L}$) para feixes submetidos a campos magn\'{e}ticos t\'{\i}picos, entre $0,\!3$\,T -- $0,\!7$\,T,  \'{e} da ordem de $10\,\mbox{GHz}$. Assim, com campos magn\'{e}ticos n\~{a}o estacion\'{a}rios, na faixa de microondas, superpostos ao campo magn\'{e}tico uniforme, e bem menos intensos,  da ordem de $10^{-3}$\,T, pode-se induzir a chamada \textit{resson\^{a}ncia de \textit{spin} do el\'{e}tron} (cuja sigla em ingl\^{e}s \'{e}} ESR), variando-se a fre\-qu\^{e}ncia at\'{e} co\-in\-cidir com a frequ\^{e}ncia de Larmor. Por resson\^{a}ncia, a raz\~{a}o giromagn\'{e}tica, o fator $g$ e o momento magn\'{e}tico, podem ser determinados com precis\~{a}o muito mai\-or do que pelo m\'{e}\-todo de deflex\~{a}o utilizado no experi\-men\-to de Stern-Gerlach.
No dom\'{\i}nio nuclear, a t\'{e}c\-ni\-ca de resson\^{a}ncia de \textit{spin} se revelou de grande utilidade na Medicina, como ser\'{a} visto na pr\'{o}\-xi\-ma se\c{c}\~{a}o.

\subsection{O \emph{spin} na formula\c{c}\~{a}o rela\-ti\-v\'{\i}s\-ti\-ca: a equa\c{c}\~{a}o de Dirac}

\begin{flushright}
\begin{minipage}{5.5cm}
\baselineskip=10pt {\small
\textit{A exist\^{e}ncia do \textit{spin}, portanto, n\~{a}o \'{e} um efeito relativ\'{\i}stico, como frequentemente se afirma, mas sim uma \textit{consequ\^{e}ncia da lineariza\c{c}\~{a}o das equa\-\c{c}\~{o}es de onda}. Isso pode ser expresso fi\-lo\-soficamente da seguinte forma: obvia\-mente, o bom Deus escreveu as equa\-\c{c}\~{o}es de campo em forma linearizada, ou seja, no caso n\~{a}o relativ\'{\i}stico, como um sistema de duas equa\c{c}\~{o}es diferenciais acopladas de primeira ordem, e ent\~{a}o acoplou o campo eletromagn\'{e}\-tico minimamente. Ele \textit{n\~{a}o} as escreveu como uma equa\c{c}\~{a}o diferencial de segunda ordem (a equa\c{c}\~{a}o de Schr\"{o}\-dinger).
 }
\smallskip

\hfill Walter Greiner
}
\end{minipage}
\end{flushright}

Com a inclus\~{a}o do \textit{spin} do el\'{e}tron na formula\c{c}\~{a}o n\~{a}o relativ\'{\i}stica da Mec\^{a}nica Qu\^{a}ntica, al\'{e}m dos efeitos Zeeman (normal e an\^{o}malo), a estrutura fina \'{e} explicada a partir da intera\c{c}\~{a}o dos momentos magn\'{e}ticos orbital e de \textit{spin} do el\'{e}tron, e se compreende que a complexidade dos multipletos resulta, principalmente, da composi\c{c}\~{a}o dos momentos angulares orbital e de \textit{spin} dos el\'{e}trons nos \'{a}tomos. No entanto, por princ\'{\i}pio, as formula\c{c}\~{o}es de Heisenberg e Schr\"{o}\-din\-ger n\~{a}o seriam v\'{a}lidas, uma vez que a Relatividade Restrita de Einstein tinha solucionado o problema da invari\^{a}ncia das leis do Eletromagnetismo em mudan\c{c}as de referenciais inerciais, mostrando que as equa\c{c}\~{o}es de Maxwell eram covariantes com rela\c{c}\~{a}o \`{a}s transforma\c{c}\~{o}es de Lorentz e, a partir de ent\~{a}o, a invari\^{a}ncia de Lorentz passou a ser um crit\'{e}rio para a aceita\c{c}\~{a}o de uma teoria fundamental da F\'{\i}sica.

A equa\c{c}\~{a}o covariante da Mec\^{a}nica Qu\^{a}ntica da part\'{\i}cula, segundo as transforma\c{c}\~{o}es de Lorentz -- a \textit{equa\c{c}\~{a}o de Dirac} --, a qual determina o comportamento do el\'{e}tron  sob a a\c{c}\~{a}o de um campo eletromagn\'{e}tico, descrito pelo potencial quadridimensional $(A_\mu) = (A_0, A_1, A_2, A_3) = (\phi/c, A_x, A_y, A_z)$, em uma regi\~{a}o do espa\c{c}o-tem\-po cujas coordenadas s\~{a}o  $ (x_\mu) \!= \! (x_0, x_1, x_2, x_3) = (ct,x,y,z)$, foi obtida por Dirac~\cite{Dirac_eletron}, em 1928, e
po\-de ser escrita como
$$   \displaystyle
\Big(\! \sum_\mu \gamma_\mu\, p_\mu - m\, c \!\Big) \, \Psi(x_\mu) \! = \! - e\, \sum_\mu \gamma_\mu\, A_\mu(x) \, \Psi(x_\mu)
$$
em que  $(\mu =0,1,2,3)$,  $(p_\mu) = (p_0, p_1, p_2, p_3)= i\, \hbar\,  \big( \partial_{x_0}, \mbox{-} \partial_{x_1},\mbox{-}  \partial_{x_2}, \mbox{-}  \partial_{x_3} \big)$ \'{e} o operador \textit{quadrimomentum}, e as matrizes ($4 \times 4$), $(\gamma_\mu)\!=\!(\gamma_0,\! \gamma_1,\! \gamma_2,\! \gamma_3)$  s\~{a}o expressas em blocos, em termos das matrizes de Pauli, como
$$ \gamma_0 = \left(\! \begin{array}{cc}
                   I_2 & 0  \\
                   0 & - I_2
                     \end{array}
           \! \right),
  \quad
 \gamma_k  = \left(\! \begin{array}{cc}
                   0 &  \sigma_k \\
                   -  \sigma_k & 0 \end{array}
            \! \right)
$$
\noindent sendo  $  (k=1,2,3),  (\sigma_1 = \sigma_x, \sigma_2 = \sigma_y, \sigma_3 = \sigma_z)$, $I_2$ a matriz identidade ($2\times 2$), e $\displaystyle \sum_\mu \gamma_\mu\, p_\mu = \gamma_0\, p_0 - \sum_k \gamma_k\, p_k = \gamma_0\, p_0 - \vec \gamma \cdot  \vec p$.

Na formula\c{c}\~{a}o relativ\'{\i}stica de Dirac, a fun\-\c{c}\~{a}o de onda ($\Psi$) \'{e} um campo espinorial de ordem 1, representado por uma matriz ($4 \times 1$), e o operador de \textit{spin} ($\vec S$) \'{e} o gerador da representa\c{c}\~{a}o espinorial quadri-dimensional do grupo das ro\-ta\-\c{c}\~{o}es, representado pela matriz ($4 \times 4$),
$$ \vec S  = \frac{\hbar}{2} \, \left(\!  \begin{array}{cc}
                                    \vec \sigma   & 0 \\
                                    0  &  \vec \sigma
                             \end{array}\! \right)
$$
em que $\vec \sigma$ \'{e} o operador de Pauli.

Um dos grandes triunfos iniciais  da teoria de Dirac foi mostrar que o valor correto para o momento magn\'{e}tico intr\'{\i}nseco do el\'{e}tron era dado pela rela\c{c}\~{a}o proposta por Goudsmit e Uhlenbeck, equa\c{c}\~{a}o~(\ref{Goud}), e adotada por Pauli,  sem a necessidade de hip\'{o}teses \textit{ad hoc} (Ap\^{e}ndice~\ref{eletro_mom_mag}). Por um longo tempo, devido ao argumento de Thomas e \`{a} dedu\c{c}\~{a}o de Dirac,  o \textit{spin} do el\'{e}tron foi considerado como um fen\^{o}meno de origem relativ\'{\i}stica, embora houvesse exce\c{c}\~{o}es proeminentes, como a de Richard Feynman~\cite{Feynman}, em 1961, como atesta Jun  Sakurai~\cite{Sakurai_a}

\begin{quotation}
\noindent\baselineskip=10pt {\small \textit{Historicamente, tudo isso foi obtido primeiro a partir do limite n\~{a}o relativ\'{\i}stico da teoria de Dirac (...). Por essa raz\~{a}o, a maioria dos livros did\'{a}ticos afirma que a rela\c{c}\~{a}o $g=2$ \'{e}
uma consequ\^{e}ncia da teoria de Dirac. Vimos, no entanto, que a re\-la\-\c{c}\~{a}o $g= 2$
segue naturalmente da te\-o\-ria n\~{a}o relativ\'{\i}stica de Schr\"{o}din\-ger-Pauli
se come\c{c}armos com o operador de energia cin\'{e}tica $H = (\vec \sigma \cdot \vec p)^2/2m$. Este ponto
foi enfatizado par\-ti\-cu\-lar\-men\-te por R. P. Feynman. }}
\end{quotation}


Em 1967, L\'{e}vy-Leblond~\cite{Leblond_1}  
 mostra que tan\-to o \textit{spin} como o momento magn\'{e}tico podem ser igualmente obtidos a partir de uma abordagem n\~{a}o relativ\'{\i}stica, e n\~{a}o apenas como limite n\~{a}o relativ\'{\i}stico da equa\c{c}\~{a}o de Dirac. Abordando o problema de forma an\'{a}loga \`{a} Dirac, procurando uma express\~{a}o linear no momentum e na energia, ele obt\'{e}m uma equa\c{c}\~{a}o covariante em rela\c{c}\~{a}o \`{a}s transforma\c{c}\~{o}es de Galileu -- a equa\c{c}\~{a}o de L\'{e}vy-Leblond --, na qual as derivadas parciais espa\c{c}o-temporais s\~{a}o de primeira ordem, que incorpora o \textit{spin} do el\'{e}tron.

\begin{equation}\label{levy-leblond}
\displaystyle
\left(\!
  \begin{array}{cc}
 \displaystyle  i\, \hbar \frac{\partial}{\partial t}  &  - c\, (\vec \sigma \cdot  \vec p) \\
    \ &  \\
    c \, (\vec \sigma \cdot \vec p ) & - 2 m c^2
  \end{array}
\! \right) \left(\! \begin{array}{c}
    \psi^+ \\
    \ \\
    \psi^-
  \end{array}
\! \right)
=  0
\end{equation}

 A partir do chamado acoplamento m\'{\i}nimo,\footnote{\, O acoplamento m\'{\i}nimo consiste em substituir o operador de \textit{momentum} na express\~{a}o $\vec \sigma \cdot \vec p$ por $(\vec p + e \vec A)$, em que $\vec A$ \'{e} o potencial vetorial, e    o operador  $i\, \hbar \frac{\partial}{\partial t}$ por $\big(i\, \hbar \frac{\partial}{\partial t} + e \phi \big)$, sendo $\phi$ o potencial escalar.} obt\'{e}m-se a \textit{equa\c{c}\~{a}o de Pauli}, a qual descreve o comportamento do el\'{e}tron em um campo eletromagn\'{e}tico, com a rela\c{c}\~{a}o correta de Goudsmit e Uhlenbeck para o momento magn\'{e}tico.

\section{O \emph{spin} 1/2 como propriedade intr\'{\i}nseca dos f\'{e}r\-mi\-ons elementares}

\begin{flushright}
\begin{minipage}{5.5cm}
\baselineskip=10pt {\small
\textit{A exist\^{e}ncia do spin e as estat\'{\i}sticas associadas a ele s\~{a}o o projeto mais sutil e engenhoso da Natureza -- sem ele, todo o universo entraria em colapso.
 }
\smallskip

\hfill Takeshi Oka
}
\end{minipage}
\end{flushright}

Uma vez estabelecido o \textit{spin} do el\'{e}tron, naturalmente, atribuiu-se tamb\'{e}m ao pr\'{o}ton, tido na \'{e}poca como uma part\'{\i}cula elementar, um momento angular de \textit{spin} 1/2, como o el\'{e}tron, e um momento magn\'{e}tico ($\mu_p$) da ordem de $1\,800$ vezes menor,
$$ \displaystyle \big(\mu_p \big)_{\mbox{\scriptsize esperado}} = \frac{\mu_e}{1\,\!837} = \frac{e \hbar}{2 m_p} = \mu_{_N} $$

\vspace*{-0.2cm}
\noindent
em que $m_p$ \'{e} a massa do pr\'{o}ton, e $\mu_N$ \'{e} o chamado \textit{magneton nuclear}.

Em 1927, David M.~Denison~\cite{Dennison} mostra que o valor de \textit{spin} 1/2 para o pr\'{o}ton era compat\'{\i}vel com os experimentos que determinaram a varia\c{c}\~{a}o com a temperatura do calor espec\'{\i}fico da mol\'{e}cula de hidrog\^{e}nio, a volume constante.

Por volta dos anos de 1930, a F\'{\i}sica j\'{a} era um ramo da Ci\^{e}ncia que havia se estendido para um grande n\'{u}mero de pa\'{\i}ses fora da Europa. A despeito da aceita\c{c}\~{a}o da comunidade cient\'{\i}fica de que o momento magn\'{e}tico do pr\'{o}ton era igual ao magneton nuclear,  Otto Stern, Otto Frisch  e  Immanuel Estermann~\cite{Stern-Frisch, Stern-Ester}, em 1933, a partir da deflex\~{a}o de feixes moleculares de hidrog\^{e}nio em campos magn\'{e}ticos,  estimam o valor do momento magn\'{e}tico do pr\'{o}ton acima do esperado,
$$ \mu_p \simeq 2,\!5 \, \mu_{_N} $$

\vspace*{-0.2cm}
\noindent
ou seja, apenas cerca de $1\,000$ vezes menor que o do el\'{e}tron, indicando que o pr\'{o}ton n\~{a}o seria uma part\'{\i}cula sem estrutura como o el\'{e}tron; uma part\'{\i}cula de Dirac.

Um valor diferente, da ordem de $3,\!25 \, \mu_{_N} $, tamb\'{e}m acima do valor esperado, foi obtido por Isidor Rabi, J. Kellogg e  J. Zacharias~\cite{Rabi_p}, em 1934, na Universidade de Columbia, tamb\'{e}m  a partir da deflex\~{a}o de feixes moleculares de hidrog\^{e}nio. Como o n\^{e}utron, descoberto em 1932, n\~{a}o tinha carga el\'{e}trica, em princ\'{\i}pio, n\~{a}o teria \textit{spin} nem momento magn\'{e}tico intr\'{\i}nseco. No entanto, ao determinarem o \textit{spin} e o momento magn\'{e}tico do d\^{e}uteron~\cite{Rabi_d}, da ordem de $0,\!77 \, \mu_{_N}$, e considerando a aditividade dos momentos magn\'{e}ticos do pr\'{o}ton e do n\^{e}utron, inferem que o n\^{e}utron, apesar da carga nula, mas sendo tamb\'{e}m uma part\'{\i}cula elementar, teria \textit{spin} 1/2, e um momento magn\'{e}tico da ordem de $\pm\,2,\!5 \, \mu_{_N}$.\footnote{\, Nessa \'{e}poca, n\~{a}o se sabia se os fatores $g$ do pr\'{o}ton e do deuteron eram positivos ou negativos. A proposta do  \textit{spin} nuclear, uma ideia j\'{a} cogitada tamb\'{e}m por Pauli, foi feita por Goudsmit e Ernst Beck, em 1927, devido a uma \textit{estrutura hiperfina} observada no espectro do bismuto. No entanto, devido ao pequen\'{\i}ssimo desdobramento das linhas espectrais, as incertezas nas determina\c{c}\~{o}es espectroc\'{o}picas eram grandes, ao contr\'{a}rio do m\'{e}todo de Stern, de deflex\~{a}o de feixes moleculares e, posteriormente, do m\'{e}todo de resson\^{a}ncia de Rabi.}

Entre 1937 e 1938, desenvolve-se uma importante t\'{e}cnica associada \`{a}s propriedades do \textit{spin} dos n\'{u}cleos. Nes\-te per\'{\i}odo, Rabi e colaboradores~\cite{Rabi_r, Rabi_1, Rabi_2}  demonstraram experimentalmente que um campo magn\'{e}tico oscilante poderia induzir transi\c{c}\~{o}es entre n\'{\i}veis associados ao estado de \textit{spin} de v\'{a}\-rios n\'{u}cleos em um cam\-po magn\'{e}tico est\'{a}tico aplicado. Este trabalho pioneiro foi realizado em feixes moleculares, utilizando um m\'{e}todo de sele\c{c}\~{a}o e detec\-\c{c}\~{a}o do estado de \textit{spin} nuclear semelhante ao desenvolvido na d\'{e}cada de 1920, por Stern e Gerlach.
 Surge, assim, o m\'{e}todo de \textit{resson\^{a}ncia magn\'{e}tica nuclear} (NMR)~\cite{Ramsey}. A partir da extra\c{c}\~{a}o de n\'{u}cleos de feixes moleculares, o grupo de Rabi determina, propriamente, o momento magn\'{e}tico do pr\'{o}ton e de v\'{a}rios n\'{u}\-cleos. Apesar da necessidade de campos magn\'{e}ticos mais intensos do que os utilizados na resson\^{a}ncia de \textit{spin} eletr\^{o}\-ni\-co, da ordem de 10~T, a frequ\^{e}ncia de resson\^{a}ncia de Larmor para a resson\^{a}ncia de \textit{spin} nuclear \'{e} bem menor, da ordem de $10^7$\,Hz, na faixa de r\'{a}dio-frequ\^{e}ncia. Enquanto a resson\^{a}ncia de \textit{spin} eletr\^{o}\-ni\-co \'{e} adequada, principalmente, para o estudo de subst\^{a}ncias paramagn\'{e}ticas, em que as mol\'{e}culas possuem uma configura\c{c}\~{a}o na qual os el\'{e}trons est\~{a}o desemparelhados, para a maioria das subst\^{a}ncias esse n\~{a}o \'{e} o caso. A resson\^{a}ncia de \textit{spin} nuclear, no entanto, permite a determina\c{c}\~{a}o da estrutura de uma ampla variedade de subst\^{a}ncias, tanto org\^{a}nicas como cristalinas. A partir de ent\~{a}o, Edward Purcell e Felix Block desenvolvem novos m\'{e}todos de medi\c{c}\~{a}o do magnetismo nuclear, que permitiram o desenvolvimento de t\'{e}cnicas apura\-das de obten\c{c}\~{a}o de imagens, hoje essenciais na Medicina, a partir da resson\^{a}n\-cia nuclear magn\'{e}tica~\cite{Nacher}.


Em 1940, com feixes de n\^{e}utrons obtidos a partir do C\'{\i}clotron da Universidade da Calif\'{o}r\-nia, em Berkeley, e utilizando tamb\'{e}m t\'{e}cnicas de resson\^{a}ncia, Luiz Alvarez e Felix Bloch~\cite{Bloch} ob\-t\^{e}m medidas diretas do \textit{spin} 1/2 e do momento mag\-n\'{e}\-ti\-co do n\^{e}utron, sem a ambiguidade do sinal. Os valores atuais dos momentos magn\'{e}ticos do pr\'{o}ton e do n\^{e}utron s\~{a}o:\footnote{\, Esses valores correspondem \`{a} metade dos valores dos respectivos fatores $g$,  $g_p = \ 5,\!585\,694\,689\,3\,(16) $\ e \ $g_n = - 3,\!826\,085\,52\,(90)$ --
CODATA (2022).}
$$
\displaystyle
\mu_p \simeq  \ 2,\!793 \, \mu_{_N} \qquad  \mbox{e} \qquad
\mu_n \simeq - 1,\!913 \, \mu_{_N}
$$

\vspace*{-0.3cm}
Durante os anos de 1950, com o in\'{\i}cio da era dos grandes aceleradores de part\'{\i}culas, o n\'{u}mero de part\'{\i}culas subat\^{o}micas e das  respectivas antipart\'{\i}culas aumenta considera\-vel\-men\-te. A primeira, a antipart\'{\i}cula do el\'{e}tron, o p\'{o}\-si\-tron, foi prevista em 1930 por Dirac~\cite{Dirac_anti}, e descoberta em 1932 por Carl Anderson~\cite{Anderson} em observa\c{c}\~{o}es de raios c\'{o}smicos. Em seguida, logo se descobre que para cada part\'{\i}cula existe uma respectiva antipart\'{\i}cula. Em 1960, o n\'{u}mero de part\'{\i}culas e antipart\'{\i}culas subat\^{o}micas j\'{a} era da ordem de centenas, em n\'{u}mero maior que o n\'{u}mero de \'{a}tomos da Tabela Pe\-ri\-\'{o}\-di\-ca. E todas tinham momento angular in\-tr\'{\i}n\-se\-co, ou seja, \textit{spin}, mas n\~{a}o s\'{o} de valor 1/2, algumas com valores inteiros ($0, 1, 2, \ldots$), e outras com outros valores semi-inteiros ($3/2, 5/2, \ldots$). Al\'{e}m disso, o \textit{spin} resultante de outras part\'{\i}culas compostas como n\'{u}cleos, \'{a}tomos e mol\'{e}culas poderia ter valores inteiros ou semi-inteiros. Com base no valor do \textit{spin} as part\'{\i}culas s\~{a}o classificadas em duas categorias:

\vspace*{-0.3cm}
\begin{description}
\item[ --] \textit{f\'{e}rmions}:  part\'{\i}culas de \textit{spin} semi-inteiro, -- como el\'{e}trons,  pr\'{o}tons e n\^{e}utrons;

\vspace*{-0.2cm}
\item[ --] \textit{b\'{o}sons}:  part\'{\i}culas de \textit{spin} inteiro, -- como \'{a}tomos de $^4${\tt He}, part\'{\i}culas $\alpha$ e d\^{e}uterons.
\end{description}

Coletivamente, os gases ideais de f\'{e}rmions est\'{a}veis e n\~{a}o relativ\'{\i}sticos obedecem \`{a} distribui\c{c}\~{a}o de Fermi-Dirac~\cite{Pauli-3, Pauli-4}, e os de b\'{o}sons massivos e n\~{a}o relativ\'{\i}sticos, \`{a} distribui\c{c}\~{a}o de Bose-Einstein~\cite{Caruso-Oguri_stat}.

A compreens\~{a}o dos valores de \textit{spin} 1/2 e dos momentos magn\'{e}ticos do pr\'{o}ton e do n\^{e}utron s\'{o} foi esclarecida com base em modelos est\'{a}ticos de constituintes, formulados em 1964, independentemente, por Murray Gell-Mann~\cite{Gell-Mann} e George Zweig~\cite{Zweig}, segundo os quais era poss\'{\i}vel classificar os \textit{h\'{a}drons}\footnote{\, Os h\'{a}drons s\~{a}o part\'{\i}culas subat\^{o}micas, como os pr\'{o}tons e os n\^{e}utrons, que est\~{a}o sujeitas \`{a}s intera\c{c}\~{o}es fortes, devido ao seu conte\'{u}do quark\^{o}nico. S\~{a}o divididos em \textit{b\'{a}rions} -- f\'{e}rmions compostos de tr\^{e}s \textit{quarks} -- e \textit{m\'{e}sons} -- b\'{o}sons compostos de dois \textit{quarks}.} at\'{e} ent\~{a}o conhecidos, embora suas premissas e os constituintes elementares postulados fossem diferentes. Segundo o modelo de \textit{quarks} de Gell-Mann, que prevaleceu, tanto o pr\'{o}ton ($p$) como o n\^{e}utron ($n$) s\~{a}o constitu\'{\i}dos por dois tipos de part\'{\i}culas elementares de \textit{spin} 1/2: o \textit{quark} $u$ de carga el\'{e}trica $\frac{2}{3} e$, e o \textit{quark} $d$ de carga el\'{e}trica $-\frac{1}{3} e$, segundo os arranjos de tr\^{e}s \textit{quarks}, $p(uud)$ e $n(udd)$.\footnote{\, No modelo inicial de classifica\c{c}\~{a}o dos h\'{a}drons s\'{o} haviam tr\^{e}s tipos, ou sabores, de \textit{quarks}; -- \textit{up} ($u$), \textit{down} ($d$) e \textit{strange} ($s$). No entanto, a descoberta de novos h\'{a}drons e argumentos te\'{o}ricos associados \`{a}s intera\c{c}\~{o}es fracas levaram \`{a} introdu\c{c}\~{a}o de mais tr\^{e}s sabores;  -- \textit{charm} ($c$), \textit{bottom} ($b$) e \textit{top} ($t$) --, e a hip\'{o}tese de mais uma simetria fundamental da natureza, associada a uma nova propriedade dos \textit{quarks}, a \textit{cor}. }

A partir da composi\c{c}\~{a}o dos \textit{spins} 1/2 desses \textit{quarks}, a raz\~{a}o obtida entre os momentos magn\'{e}ticos~\cite{Caruso_QCD},
$$ \displaystyle \left(\frac{\mu_n}{\mu_p}\right)_{\mbox{\scriptsize  teor}} \hspace*{-0.5cm} = - \frac{2}{3} \simeq 0,\!67$$

\vspace*{-0.2cm}
\noindent
comparada com o valor experimental da \'{e}poca,
$$ \displaystyle \left(\frac{\mu_n}{\mu_p}\right)_{\mbox{\scriptsize  exp}}\hspace*{-0.3cm} = - \frac{2}{3} \simeq 0,\!68$$

\vspace*{-0.3cm}
\noindent foi fundamental para o sucesso do modelo.

\vspace*{-0.2cm}
A ampla aceita\c{c}\~{a}o do modelo dos \textit{quarks} a\-con\-tece ap\'{o}s o experimento de espalhamento de e\-l\'{e}\-trons por pr\'{o}tons, em 1969, no acelerador linear de Stanford (SLAC), por grupos do MIT\footnote{\, Instituto de Tecnologia de Massachusetts.} e do SLAC, e uma s\'{e}rie de outros experimentos nos anos de 1970, em particular o espalhamento de neutrinos por n\'{u}cleons (pr\'{o}tons ou n\^{e}utrons).

\vspace*{-0.2cm}
Ap\'{o}s a consolida\c{c}\~{a}o do Modelo Padr\~{a}o da F\'{\i}sica de Part\'{\i}culas, nos anos de 1970, o \textit{spin} 1/2 \'{e} reconhecido como uma propriedade  in\-tr\'{\i}n\-se\-ca atribu\'{\i}da aos f\'{e}rmions elementares; os l\'{e}ptons e \textit{quarks} -- os constituintes fundamentais da mat\'{e}ria (Tabela~\ref{modelo_padrao}).\footnote{\, Al\'{e}m dos f\'{e}rmions elementares e de suas respectivas antipart\'{\i}culas, o Modelo Padr\~{a}o cont\'{e}m os b\'{o}sons de \textit{spin} 1, mediadores das intera\c{c}\~{o}es fundamentais: o f\'{o}ton -- mediador da intera\c{c}\~{a}o eletromagn\'{e}tica; o $W^+$, o $W^-$ e o $Z$ -- mediadores das intera\c{c}\~{o}es fracas; oito tipos de gl\'{u}ons -- mediadores das intera\c{c}\~{o}es fortes, e o b\'{o}son de Higgs, de \textit{spin} 0. No dom\'{\i}nio subat\^{o}mico, a intera\c{c}\~{a}o gravitacional n\~{a}o \'{e} levada em conta.}

\begin{table}[hbtp]
\caption{\normalsize Os f\'{e}rmions elementares do Modelo Padr\~{a}o (\textit{spin} 1/2)}
\renewcommand{\arraystretch}{1.}
\begin{center}
\begin{tabular}{ l c  c }
    \hspace{0.1cm} l\'{e}ptons    &  carga($e$) & massa (Mev/c$^2$)     \\
\hline
$\begin{array}{l}
\mbox{\scriptsize neutrino} \vspace*{-0.2cm} \\
 \mbox{\scriptsize do el\'{e}tron}
 \end{array}$ \hspace{-0.3cm}
($\nu_e$)       &  ~0              &  $<  10^{-6}$ \\
\hspace{0.08cm} \mbox{\scriptsize el\'{e}tron} ($e$)   &  -1    &  $0,\!511$   \\
$\begin{array}{l}
\mbox{\scriptsize neutrino} \vspace*{-0.2cm} \\
 \mbox{\scriptsize do m\'{u}on}
 \end{array}$ \hspace{-0.3cm}
($\nu_\mu$)  &  ~0              &  $<  10^{-6}$ \\
\hspace{0.08cm} \mbox{\scriptsize m\'{u}on} ($\mu$)    &  -1  &  $105,\!7$ \\
$\begin{array}{l}
\mbox{\scriptsize neutrino} \vspace*{-0.2cm} \\
 \mbox{\scriptsize do tau}
 \end{array}$ \hspace{-0.3cm}
($\nu_\tau$)   &  ~0              &  $< 10^{-6}$ \\
\hspace{0.08cm} \mbox{\scriptsize tau} ($\tau$)          &  -1  & $1\, 776,\!9 $
\end{tabular}\\
\vspace*{0.2cm}
\begin{tabular}{ l c  c   }
 \textit{quarks}  &  carga($e$) & massa (Gev/c$^2$) \\
\hline
 \mbox{\scriptsize \textit{up}} ($u$)    & ~2/3              &  $2,\!16 \times 10^{-3}$  \\
  \mbox{\scriptsize \textit{down}}  ($d$)          & -1/3           &  $4,\!70 \times 10^{-3}$ \\
   \mbox{\scriptsize \textit{charm}} ($c$)                   & ~2/3              &  $1,\!27$  \\
      \mbox{\scriptsize \textit{strange}}  ($s$)                   & -1/3              &  $93,\!5 \times 10^{-3}$ \\
       \mbox{\scriptsize \textit{top}} ($t$)                   & ~2/3              &  $172,\!6 $ \\
         \mbox{\scriptsize \textit{bottom}} ($b$)                   & -1/3              &  $4,\!18 $
\end{tabular}
\end{center}
\label{modelo_padrao}
\end{table}
\renewcommand{\arraystretch}{1.}

Em princ\'{\i}pio, o \textit{spin} das part\'{\i}culas sub\-a\-t\^{o}\-mi\-cas n\~{a}o elementares -- os h\'{a}drons --, resultaria da composi\c{c}\~{a}o de seus \textit{quarks} constituintes. No entanto, o  experimento de espalhamento de m\'{u}ons por pr\'{o}tons~\cite{Ashman}, realizado no CERN,\footnote{\, Laborat\'{o}rio Europeu de F\'{\i}sica de Part\'{\i}culas, em Genebra, Su\'{\i}\c{c}a.} em 1987, mostrou que a contribui\c{c}\~{a}o do \textit{quarks} para o \textit{spin} do pr\'{o}ton era muito menor que o esperado. Esse resultado e outros similares, obtidos posteriormente, conhecidos como ``a crise do \textit{spin} do pr\'{o}ton'', ainda permanecem como um dos problemas n\~{a}o explicados pela F\'{\i}sica Te\'{o}rica.

Do mesmo modo que os experimentos de espalhamentos de el\'{e}trons por pr\'{o}tons mostraram que os \textit{quarks} s\~{a}o respons\'{a}veis por apenas cerca de 50\% do \textit{momentum} do pr\'{o}ton, sendo o restante atribu\'{\i}do aos gl\'{u}ons, uma parte dos pesquisadores consideram que os gl\'{u}ons sejam respons\'{a}veis tamb\'{e}m por uma grande parcela do \textit{spin}. Ainda h\'{a} outros que consideram que uma grande parcela deve-se ao momento angular orbital dos \textit{quarks} no interior do pr\'{o}ton~\cite{Leader_1}.

\section{Considera\c{c}\~{o}es finais}

Para al\'{e}m das novas possibilidades experimentais e te\'{o}ricas abertas a partir da descoberta do \textit{spin}, como foi discutido ao longo do artigo, cabe, por fim, destacar seu papel na afirma\c{c}\~{a}o do atomismo. De fato, do ponto de vista epistemol\'{o}gico, a for\c{c}a da concep\c{c}\~{a}o atom\'{\i}stica da mat\'{e}ria e seu papel basilar na constru\c{c}\~{a}o do pensamento cient\'{\i}fico moderno pode ser inferida a partir do seguinte coment\'{a}rio de Feynman:
\index{nome Feynman (1918-1988)! Richard}

\vspace*{-0.2cm}
\begin{quotation}
\noindent\baselineskip=10pt {\small
\textit{Se, em algum cataclismo, todo o conhecimento cient\'{\i}fico fosse destru\'{\i}do e somente uma senten\c{c}a fosse transmitida para as pr\'{o}ximas gera\c{c}\~{o}es de criaturas, \, que\ enunciado conteria\ mais informa\c{c}\~{a}o em menos pala\-vras? Acredito que seja a {\em hip\'{o}tese at\^{o}mica} {\rm [...]} de que {\em todas as coisas s\~{a}o feitas de \'{a}tomos} {\rm [...]}. Nessa \'{u}nica senten\c{c}a, voc\^{e} ver\'{a}, existe uma {\em enorme} quantidade de informa\c{c}\~{o}es sobre o Mun\-do.}}
\end{quotation}

\vspace*{-0.2cm}
Talvez o maior \'{\i}cone dessa conquista seja a Tabela Peri\'{o}dica de Mendeleiev, express\~{a}o ine\-g\'{a}\-vel de uma grande s\'{\i}ntese resultante da acumula\c{c}\~{a}o de observa\c{c}\~{o}es de regularidades f\'{\i}sico-qu\'{\i}micas macrosc\'{o}picas associadas aos elementos fundamentais (os \'{a}tomos) da Natureza, reflexos de uma simetria maior, ainda escondida \`{a} \'{e}poca do qu\'{\i}mico e f\'{\i}sico russo. A partir da descoberta do el\'{e}tron, coube \`{a} F\'{\i}sica prover modelos at\^{o}micos que, gradativamente, contribu\'{\i}am para elucidar a origem dessa fant\'{a}stica simetria, at\'{e} que se chegou a uma nova teoria para o microcosmo -- a \textit{Mec\^{a}nica Qu\^{a}ntica}. Em particular, a chave para a resposta a essa indaga\c{c}\~{a}o foi a introdu\c{c}\~{a}o do conceito de \textit{spin}, o qual passa a ser efetivamente compreendido no \^{a}mbito dessa nova teoria torna-se essencial para se compreendera realidade e as liga\c{c}\~{o}es qu\'{\i}micas, determinando quantos el\'{e}trons podem ser compartilhados numa liga\c{c}\~{a}o. A pr\'{o}pria distribui\c{c}\~{a}o eletr\^{o}nica de qualquer \'{a}tomo \'{e} ditada pelo \textit{prin\-c\'{\i}\-pio de exclus\~{a}o de Pauli}.

Tem-se, ent\~{a}o, um quadro te\'{o}rico coerente no qual o \textit{princ\'{\i}pio de exclus\~{a}o} e a antissimetria da fun\c{c}\~{a}o de ondas de dois f\'{e}rmions coexistem. \'{E} um equ\'{\i}voco afirmar que o princ\'{\i}pio de Pauli´pode ser provado teoricamente a partir da imposi\c{c}\~{a}o dessa antissimetria. Trata-se apenas da troca de uma hip\'{o}tese por outra. Se assim n\~{a}o fosse, n\~{a}o faria mais sentido continuar usando o termo ``princ\'{\i}pio''.

Por outro lado, o \textit{spin} traz consigo uma novidade sem paralelo na F\'{\i}sica Cl\'{a}ssica. \'{E} a ideia de que as part\'{\i}culas se caracterizam, principalmente, pela natureza de seu \textit{spin}, em dois grandes grupos: b\'{o}sons e f\'{e}rmions. Conforme foi recordado, Dirac havia estabelecido, em 1926, que a fun\c{c}\~{a}o de onda de um sistema de duas part\'{\i}culas deve ser sim\'{e}trica pela troca das par\-t\'{\i}\-cu\-las, no caso de b\'{o}sons, e antissim\'{e}trica, no caso de f\'{e}rmions. Al\'{e}m disso, uma part\'{\i}cula de \textit{spin} 0 \'{e} descrita por um campo escalar, uma de \textit{spin} 1/2, por um campo espinorial, enquanto outra de \textit{spin} 1, por um campo vetorial. Desse modo, todas as rea\c{c}\~{o}es e decaimentos em F\'{\i}sica de Part\'{\i}culas dependem, em \'{u}ltima an\'{a}\-li\-se, dos valores dos spins das part\'{\i}culas envolvidas. Como consequ\^{e}ncia, uma s\'{e}rie de regras de sele\-\c{c}\~{a}o se imp\~{o}e devido ao \textit{spin} e \`{a} conserva\c{c}\~{a}o do momento angular total.


Por fim, deve-se destacar a relev\^{a}ncia do papel do \textit{spin} em testes mais precisos do Modelo Padr\~{a}o, em processos nos quais se procura identificar o estado de todas as part\'{\i}culas que participam de uma colis\~{a}o ou decaimento em altas energias -- os \textit{processos exclusivos}. No entanto, na maioria dos experimentos cujos resultados s\~{a}o compat\'{\i}veis com as previs\~{o}es dos c\'{a}lculos perturbativos da teoria das intera\c{c}\~{o}es fortes,  a \textit{Cromodin\^{a}mica Qu\^{a}ntica} (QCD), identi\-fi\-cam-se apenas uma ou algumas poucas part\'{\i}culas ao final dos chamados \textit{processos inclusivos}. Em um processo inclusivo de colis\~{a}o h\'{a}dron-h\'{a}dron em altas energias, considerando as intera\c{c}\~{o}es fortes entre os constituintes hadr\^{o}nicos (\textit{quarks}), a probabilidade associada a um dado estado final, ou a \textit{se\c{c}\~{a}o de choque do processo}, \'{e} determinada pela \textit{soma incoerente} das se\c{c}\~{o}es de choque dos subprocessos elementares entre os \textit{quarks}. Essa soma incoerente resulta de uma m\'{e}dia sobre os \textit{momenta} e \textit{spins} de todos os constituintes  dos h\'{a}drons envolvidos. No caso de um processo exclusivo, antes de se calcular a \textit{se\c{c}\~{a}o de choque do processo}, deve-se somar as amplitudes de probabilidade de todos os pos\-s\'{\i}\-veis subprocessos elementares, que dependem, em \'{u}l\-ti\-mo caso, do \textit{spin} das part\'{\i}culas envolvidas. Como consequ\^{e}ncia, h\'{a} a manifesta\c{c}\~{a}o de efeitos de interfer\^{e}ncia de estados de \textit{spin}.
Espera-se, portanto, que a realiza\c{c}\~{a}o de futuros experimentos envolvendo processos exclusivos, al\'{e}m de testes da QCD, em regimes perturbativos, de altas transfer\^{e}ncias de \textit{quadrimomentum}, e n\~{a}o perturbativos, de baixas transfer\^{e}ncias de \textit{quadrimomentum}, possibilite uma melhor compreens\~{a}o do papel dos gl\'{u}ons na composi\c{c}\~{a}o dos \textit{spins} dos h\'{a}drons.

Al\'{e}m das fontes originais apresentadas no texto, outros tantos aspectos envolvidos na e\-vo\-lu\-\c{c}\~{a}o do conceito do \textit{spin}, podem ser encontrados nas refer\^{e}ncias~\cite{Greiner, Landau, Feynman_lect, Auletta, Susskind, Frohlich}.

\appendix

\section{A precess\~{a}o de Larmor}
\label{Larmor_classica}

Segundo o Eletromagnetismo Cl\'{a}ssico, o momento dipolar magn\'{e}tico ($\vec \mu$) e o momento angular ($\vec L$) de uma part\'{\i}cula de massa $m$ e carga el\'{e}trica $q$, segundo o mesmo ponto de um referencial inercial, est\~{a}o relacionados no SI por
$$ \vec \mu \ = \  \displaystyle \frac{q}{2m} \, \vec L $$
ou seja, ambos s\~{a}o colineares, e t\^{e}m o mesmo sentido se a carga da part\'{\i}cula \'{e} positiva, e sen\-tidos contr\'{a}rios, se a carga \'{e} negativa.

\begin{figure}[htbp]
\centerline{\includegraphics[width=4.2cm]{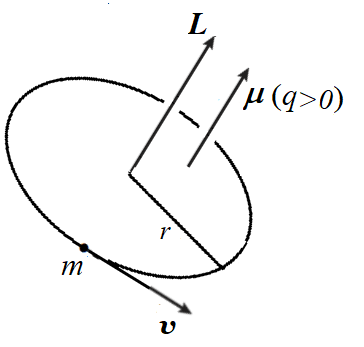}}
\caption{\baselineskip=8pt{\small Diagrama destacando algumas grandezas f\'{\i}sicas associadas a uma} part\'{\i}cula de massa $m$ e carga el\'{e}trica $q$ em movimento circular uniforme.}
\label{carga_circ}
\end{figure}

Se a part\'{\i}cula est\'{a} sob a a\c{c}\~{a}o de uma for\c{c}a central, em movimento circular uniforme, o momento angular \'{e} uma constante do movimento, e segundo a origem da for\c{c}a, perpendicular ao plano da \'{o}rbita (Figura~\ref{carga_circ}).

Se a part\'{\i}cula sofre tamb\'{e}m a a\c{c}\~{a}o de um campo magn\'{e}tico uniforme ($\vec B$), estar\'{a} sujeita igualmente a um torque dado por $\vec \mu \times \vec B$, o qual, segundo a Mec\^{a}nica Cl\'{a}ssica, \'{e} igual a varia\c{c}\~{a}o temporal do momento angular,
\begin{equation}\label{larmor_dipolo}
 \displaystyle \frac{{\rm d} \vec L}{{\rm d} t} = \vec \mu \times \vec B = \left( \frac{q}{2 m} \right) \, \vec L \times \vec B
 \end{equation}

A equa\c{c}\~{a}o~(\ref{larmor_dipolo}) indica que apenas a dire\c{c}\~{a}o do momento angular \'{e} alterada, a magnitude e o \^{a}ngulo ($\theta$) que faz com a dire\c{c}\~{a}o do campo permanecem constantes ao longo do tempo.

\begin{figure}[htbp]
\centerline{\includegraphics[width=3.8cm]{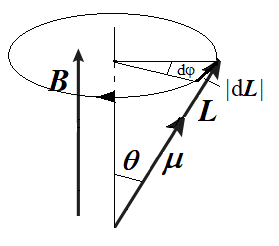}}
\caption{\baselineskip=8pt{\small Precess\~{a}o de Larmor.}}
\label{larmor_precessa}
\end{figure}

Dessa forma, o torque decorrente do campo magn\'{e}tico n\~{a}o causa o alinhamento do momento angular da part\'{\i}cula com o campo. De modo an\'{a}logo \`{a} precess\~{a}o de um pi\~{a}o no campo gravitacional terrestre,  a reta ao longo da dire\c{c}\~{a}o dos momentos angular e dipolar gira de um \^{a}ngulo ($\varphi$) em torno do eixo definido pela dire\c{c}\~{a}o do campo magn\'{e}tico. Nesse contexto, o comportamento \'{e} denominado precess\~{a}o de Larmor e, de acordo com a Figura~~\ref{larmor_precessa},
\begin{equation}\label{larmor_precessao}
\displaystyle \big| {\rm d}\vec  L \big| \! =\! \big( L \, {\rm sen}\, \theta \big) \, {\rm d} \varphi \quad \Rightarrow \quad \left| \frac{{\rm d} \vec L}{{\rm d} t} \right|\! =\! L \, {\rm sen}\, \theta \, \frac{{\rm d} \varphi}{{\rm d} t}
 \end{equation}
a varia\c{c}\~{a}o temporal do \^{a}ngulo $\varphi$, denominada frequ\^{e}ncia de Larmor ($\omega_{_L}$), \'{e} dada por
$$\displaystyle \frac{{\rm d} \varphi}{{\rm d} t} = \omega_{_L} =  \left( \frac{|q|}{2 m } \right) B  = \gamma B$$
em que $\displaystyle \gamma = \frac{|q|}{2m}$ \'{e} a \textit{raz\~{a}o giromagn\'{e}tica} da part\'{\i}cula.

Segundo a equa\c{c}\~{a}o~(\ref{larmor_precessao}), independentemente da orienta\c{c}\~{a}o do plano da \'{o}rbita com rela\c{c}\~{a}o ao campo, enquanto a componente do momento angular perpendicular ao campo gira em movimento circular uniforme com a frequ\^{e}ncia de Larmor, a componente na dire\c{c}\~{a}o do campo se mant\'{e}m constante.

\begin{figure}[htbp]
\centerline{\includegraphics[width=4.5cm]{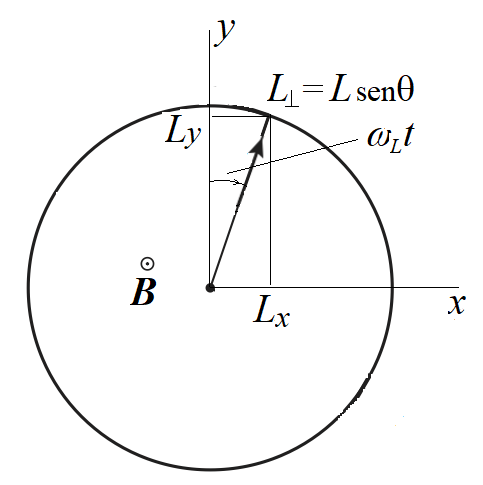}}
\caption{\baselineskip=8pt{\small Diagrama das oscila\c{c}\~{o}es das componentes do momento angular no plano $xy$, perpendicular ao campo magn\'{e}tico.}}
\label{larmor_comp}
\end{figure}

Para um sistema de coordenadas cartesianas $(x,y,z)$, e um campo magn\'{e}tico no sentido positivo da dire\c{c}\~{a}o $z$ (Figura~\ref{larmor_comp}), as componentes $L_x$ e $L_y$ do momento oscilam no plano $xy$, respectivamente, ao longo das dire\c{c}\~{o}es $x$ e $y$, enquanto a componente $L_z$ \'{e} constante, e $L\,{\rm sen}\,\theta = L_\perp$ gira com velocidade angular $\omega_L$. Com uma escolha conveniente das condi\c{c}\~{o}es iniciais, pode-se escrever
$$ \left\{
\begin{array}{l}
 \displaystyle L_x = \big(L\,{\rm sen}\,\theta \big) \, {\rm sen}\,   \omega_{_L} t \\
 \ \\
\displaystyle L_y = \big(L\,{\rm sen}\,\theta \big)  \cos \omega_{_L} t \\
 \ \\
 L_z = \big(L\,\cos \theta \big)
 \end{array}
 \right.
$$

Quaisquer que sejam as condi\c{c}\~{o}es iniciais, se o momento angular da part\'{\i}cula n\~{a}o tem a mes\-ma dire\c{c}\~{a}o do campo magn\'{e}tico aplicado, suas componentes no plano perpendicular ao cam\-po oscilam com a frequ\^{e}ncia de Larmor.


\section{O momento magn\'{e}tico do el\'{e}tron}
\label{eletro_mom_mag}

\vspace*{-0.3cm}
A equa\c{c}\~{a}o de Dirac pode ser escrita como
\begin{eqnarray*}
  \sum_\mu  \gamma_\mu \, \big( p_\mu + e \, A_\mu \big)\, \Psi  &=&  m\, c \,\Psi \\
  \Big[ \gamma_0 \, \big( p_0  + e A_0 ) - \vec \gamma \cdot \big( \vec p + e \vec A \big) \Big]\, \Psi  &=&  m\, c \,\Psi
\end{eqnarray*}
ou, na forma matricial, como\footnote{\, $\Psi$, o chamado \textit{espinor de Dirac}, \'{e}, comumente, expresso em termos dos espinores de Pauli, $\psi_A$ e $\psi_B$.}
$$
\begin{array}{l}
\displaystyle
\left(\!
  \begin{array}{cc}
    (p_0\, c + e \, \phi) & - c\, \vec \sigma \cdot ( \vec p + e \vec A ) \\
    \ &  \\
    c \,\vec \sigma \cdot ( \vec p + e \vec A ) & - ( p_0\, c + e \, \phi)
r  \end{array}
\! \right) \left(\! \begin{array}{c}
    \psi_A \\
    \ \\
    \psi_B
  \end{array}
\! \right)
 =  \\
 \ \\
 \qquad  =
  \left(\!
  \begin{array}{cc}
    m c^2   & 0 \\
    \ & \\
    0 & m c^2
  \end{array}
\! \right) \left(\! \begin{array}{c}
    \psi_A \\
    \ \\
    \psi_B
  \end{array}
\! \right)
\end{array}
$$

Explicitando as equa\c{c}\~{o}es, obt\'{e}m-se
$$ \displaystyle
\left\{ \!
   \begin{array}{l }
    (p_0\, c + e\, \phi)\, \psi_A  - c \,\vec \sigma \cdot ( \vec p + e \vec A )\, \psi_B = m c^2 \, \psi_A \\
    \   \\
    c\, \vec \sigma \cdot ( \vec p + e \vec A )\, \psi_A - ( p_0\, c + e \, \phi) \, \psi_B = m c^2\, \psi_B
  \end{array}
 \right.
$$

\vspace*{-0.3cm}
No limite n\~{a}o relativ\'{\i}stico e de campo fraco, $\displaystyle  \psi \sim e^{-i m c^2 t/\hbar}$\  e \ $ p_0\, c \,\psi  \approx m c^2  \, \psi$, pode-se ainda reescrever as duas \'{u}ltimas equa\c{c}\~{o}es, na ordem inversa, como
\begin{equation}\label{pequena}
 \displaystyle \psi_B =  \frac{1}{2mc} \, \vec \sigma \cdot \big( \vec p + e \vec A \big) \psi_A
\end{equation}
e
{\small \begin{equation}\label{grande}
\displaystyle
\left\{ (p_0\, c + e\, \phi)\, - \frac{1}{2m} \, \Big[\vec \sigma \cdot \big( \vec p + e \vec A \big) \Big]^2 \right\} \psi_A  = m c^2 \, \psi_A
\end{equation}}

\vspace*{-0.1cm}
Levando-se em conta as seguintes rela\c{c}\~{o}es~\cite{Sakurai_b, Caruso-Oguri_FM}:
{\small $$ \left\{ \!
\begin{array}{l}
\big(\vec \sigma \cdot \vec a\big)\, \big(\vec \sigma \cdot \vec b\big) = \big(\vec a \cdot \vec b\big) + i\, \vec \sigma \cdot \big(\vec a \times \vec b\big) \\
\ \\
i \vec \sigma \cdot \big(\vec p \times \vec A  + \vec A \times \vec p \big) = \hbar \, \vec \sigma \cdot \big( \vec \nabla \times \vec A \big) = \hbar \, \vec \sigma \cdot \vec B
\end{array}
\right.
$$}
em que $\vec p = - i \,\hbar \, \vec \nabla$\ e \ $ \vec \nabla \times \vec A  = \vec B$, o termo quadr\'{a}tico torna-se
$$\left\{\begin{array}{l} \Big[\vec \sigma \cdot \big( \vec p + e \vec A \big) \Big]^2 = \underbrace{\big( \vec \sigma \cdot \vec p \big)^2}_{p^2} + \\
\qquad + \ e \, \Big[ \underbrace{
\big(\vec \sigma \cdot \vec p\big)\, \big(\vec \sigma \cdot \vec A\big) + \big(\vec \sigma \cdot \vec A \big)\, \big(\vec \sigma \cdot \vec p\big)}_{\big(\vec p  \cdot \vec A  + \vec A \cdot \vec p \big) + \hbar \, \vec  \sigma \cdot \vec B} \Big] + \\
\qquad + \ e^2 \, \underbrace{\big(\vec \sigma \cdot \vec A\big)^2}_{A^2}
\end{array}\right.$$

\vspace*{-0.3cm}
Uma vez que $m c^2$ \'{e} uma constante, e o fator $1/c$ na equa\c{c}\~{a}o~(\ref{pequena}) indica que a componente $\psi_B$ pode ser desprezada face \`{a} componente $\psi_A$, no limite n\~{a}o relativ\'{\i}stico, a equa\c{c}\~{a}o que descreve o comportamento do el\'{e}tron sob a a\c{c}\~{a}o de um campo eletromagn\'{e}tico \'{e} dada pela chamada \textit{equa\c{c}\~{a}o de Pauli}~\cite{Pauli_spin}:
{\small $$ \displaystyle i\, \hbar \, \frac{\partial}{\partial t} \psi \! =\! \left[ \frac{1}{2m} \, \big( \vec p + e \, \vec A \big)^2 \!  -  e\, \phi + \left( \frac{e \hbar}{2m}\right) \vec \sigma \cdot \vec B \right] \psi $$}

\vspace*{-0.3cm}
Identificando o termo de energia em um cam\-po magn\'{e}tico,
$$ - \vec \mu \cdot \vec B = \Big(\underbrace{ \frac{e \hbar}{2m}}_{\mu_{_B}} \Big) \vec \sigma \cdot \vec B$$
obt\'{e}m-se a rela\c{c}\~{a}o de Goudsmit e Uhlenbeck para o momento magn\'{e}tico do el\'{e}tron,
$$  \vec \mu = - \mu_{_B} \vec \sigma = g_e \left( \frac{e}{2m} \right) \vec S \quad (g_e = -2)$$




\end{document}